\documentstyle[12pt,epsf]{article}

\def\Re{\mbox{Re}}
\def\Im{\mbox{Im}}
\def\di12r{$\Delta I=1/2$ rule}

\def\openone{{\displaystyle\mathbf 1}}
\def\tableline{\hline}

\date{}

\begin{document}

\title{Matrix elements relevant for $\Delta I=1/2$ rule 
and $\varepsilon '/\varepsilon$ from Lattice QCD with staggered fermions}

\author{D. Pekurovsky and G. Kilcup \\ \\
Department of Physics, \\
the Ohio State University, \\
174 W. 18th Ave., Columbus OH 43210, USA}       

\begin{titlepage}

\maketitle

\begin{abstract}
We perform a study of matrix elements relevant for the $\Delta I=1/2$ rule
and the direct CP-violation parameter $\varepsilon '/\varepsilon$ 
from first principles
by computer simulation in Lattice QCD. We use staggered (Kogut-Susskind)
fermions, and employ the chiral perturbation theory method 
for studying $K^0 \to \pi\pi$ decays. 
Having obtained a reasonable statistical accuracy, we observe an 
enhancement of the $\Delta I=1/2$ 
amplitude, consistent with experiment within our large systematic
errors. Finite volume and quenching effects have been studied
and were found small compared to noise. The estimates of
$\varepsilon '/\varepsilon$ are hindered by large uncertainties
associated with operator matching. 
In this paper we explain the simulation method, present the results and 
address the systematic uncertainties. 

\end{abstract}

\thispagestyle{empty}

\end{titlepage}

\section{Introduction}

In those areas of particle phenomenology which require addressing
non-perturbative effects, Lattice QCD plays an increasingly
significant role, being a first-principles method. The rapid advances in
computational performance as well as algorithmic techniques
are allowing for better control of various errors associated
with lattice calculations. 

In this paper we address the phenomenology of $K^0\to\pi\pi$ decays.
One of the long-standing puzzles is the ``$\Delta I=1/2$ rule'',
which is the observation that the transition channel with isospin changing
by 1/2 is enhanced 22 times with respect to transitions
with isospin changing by 3/2. 
The strong interactions are essential for
explaining this effect within the Standard Model.
Since the energy scales involved in these decays are rather small,
computations in quantum chromodynamics (QCD) have to be done
using a non-perturbative method such as Lattice QCD. Namely,
Lattice QCD is used to calculate hadronic matrix elements of 
the operators appearing in the effective weak Hamiltonian. 

There have been so far several other attempts to study matrix elements of the 
operators relevant for $\Delta I=1/2$ rule on the 
lattice~\cite{KilcupSharpe,BernardSoni,MartinelliMaiani},
but they fell short of desired accuracy.
In addition, several groups~\cite{Twopi,BernardSoniTwopi} have studied matrix
elements $\langle \pi^+\pi^0|O_i|K^+\rangle$, which describe
only $\Delta I=3/2$, not $\Delta I=1/2$ transition.
In the present simulation, the statistics is finally under control
for $\Delta I=1/2$ amplitude. 

Our main work is in calculating matrix elements 
$\langle\pi^+ |O_i|K^+\rangle$ and $\langle 0|O_i|K^0\rangle$ for
all basis operators (introduced in Sec.~\ref{sec:framework}). 
This is enough to recover matrix elements
$\langle\pi\pi|O_i|K^0\rangle$ using chiral perturbation theory
in the lowest order, although this procedure suffers from 
uncertainties arising from ignoring higher orders (in particular, 
final state interactions). The latter matrix elements are an essential
part of the phenomenological expressions for $\Delta I=1/2$ 
and \mbox{$\Delta I=3/2$} amplitudes, as well as $\varepsilon '/\varepsilon$.
The ratio of the amplitudes computed in this way
confirms significant enhancement of $\Delta I=1/2$ channel,
although systematic uncertainties preclude a definite answer.

In addition, we address a related issue of 
$\varepsilon '/\varepsilon$ -- the direct 
CP-violation parameter in the neutral kaon system.
As of the day of writing, the experimental data are somewhat 
ambiguous about this parameter: the group at CERN (NA48)~\cite{CERN} 
reports $\Re (\varepsilon '/\varepsilon) = \mbox{
$(23 \pm 7) \times 10^{-4},$}$ 
while the Fermilab group \mbox{(E731)}~\cite{Fermilab} has found 
$\Re (\varepsilon '/\varepsilon) = \mbox{$(7.4 \pm 6.0) \times 10^{-4}$.}$ 
There is a hope that the discrepancy between the two reports will soon
be removed in a new generation of experiments.

On the theoretical side, the progress in estimating 
$\varepsilon '/\varepsilon$ in the Standard
Model is largely slowed down by the unknown matrix elements~\cite{buras} 
of the appropriate operators.
The previous attempts~\cite{KilcupSharpe,BernardSoni,MartinelliMaiani} 
to compute them on the lattice did not take
into account operator matching. In this work we repeat this calculation
with better statistics and better investigation of systematic 
uncertainties. We are using perturbative operator matching. In some cases
it does not work, so we explore alternatives and come up with a
partially non-perturbative renormalization procedure. The associated
errors are estimated to be large. This is currently the biggest stumbling 
block in computing $\varepsilon '/\varepsilon$. 

The paper is structured as follows. In the Section~\ref{sec:Framework}
we show the context of our calculations, define the quantities
we are looking after and discuss a number of theoretical points
relevant for the calculation. Section~\ref{sec:lattice details}
discusses issues pertaining to the lattice simulation. 
In Section~\ref{sec:di12} we present the results and discuss systematic errors
for $\Delta I=1/2$ rule amplitudes.
In Section~\ref{sec:pert} we explain how the operator matching problem 
together with other systematic errors preclude a reliable calculation of 
$\varepsilon '/\varepsilon$, and give our best estimates
for this quantity in Section~\ref{sec:epsp_res}.
Section~\ref{sec:conclusion} contains the conclusion. In the Appendix
we give details about the quark operators and sources, and
provide explicit expressions for all contractions and matrix
elements for reference purposes.

\section{Theoretical framework}
\label{sec:Framework}

\subsection{Framework and definitions}
\label{sec:framework}

The standard approach to describe the problems in question is to
use the Operator Product Expansion at the $M_W$ scale and use the
Renormalization Group equations to translate the effective weak
theory to more 
convenient scales ($\mu \sim$~2--4~GeV). At these scales the effective 
Hamiltonian for $K\to\pi\pi$ decays is the following linear 
superposition~\cite{buras}:
\begin{equation}
H_{\mathrm W}^{\mathrm eff} = 
\frac{G_F}{\sqrt{2}} V_{ud}\,V^*_{us} \sum_{i=1}^{10} \Bigl[
z_i(\mu) + \tau y_i(\mu) \Bigr] O_i (\mu) 
 \, , 
\end{equation}
where $z_i$ and $y_i$ 
are Wilson coefficients (currently known at two-loop order), 
$\tau \equiv - V_{td}V_{ts}^{*}/V_{ud} V_{us}^{*}$, 
and $O_i$ are basis of four-fermions operators defined as follows:
\begin{eqnarray}
\label{eq:ops1}
O_1 & = & (\bar{s}_\alpha \gamma_\mu (1-\gamma_5) u_\beta )
(\bar{u}_\beta \gamma^\mu (1-\gamma_5)d_\alpha )  \\
O_2 & = & (\bar{s}_\alpha \gamma_\mu (1-\gamma_5)u_\alpha)
(\bar{u}_\beta \gamma^\mu (1-\gamma_5)d_\beta )  \\
\label{eq:ops3}
O_3 & = & (\bar{s}_\alpha \gamma_\mu (1-\gamma_5)d_\alpha )
\sum_q(\bar{q}_\beta \gamma^\mu (1-\gamma_5)q_\beta ) \\
O_4 & = & (\bar{s}_\alpha \gamma_\mu (1-\gamma_5)d_\beta )
\sum_q(\bar{q}_\beta \gamma^\mu (1-\gamma_5)q_\alpha ) \\
O_5 & = & (\bar{s}_\alpha \gamma_\mu (1-\gamma_5)d_\alpha )
\sum_q(\bar{q}_\beta \gamma^\mu (1+\gamma_5)q_\beta )  \\
O_6 & = & (\bar{s}_\alpha \gamma_\mu (1-\gamma_5)d_\beta )
\sum_q(\bar{q}_\beta \gamma^\mu (1+\gamma_5)q_\alpha )  \\
O_7 & = & \frac{3}{2}(\bar{s}_\alpha \gamma_\mu (1-\gamma_5)d_\alpha )
\sum_q e_q (\bar{q}_\beta \gamma^\mu (1+\gamma_5)q_\beta ) \\
O_8 & = & \frac{3}{2}(\bar{s}_\alpha \gamma_\mu (1-\gamma_5)d_\beta )
\sum_q e_q (\bar{q}_\beta \gamma^\mu (1+\gamma_5)q_\alpha ) \\
O_9 & = & \frac{3}{2}(\bar{s}_\alpha \gamma_\mu (1-\gamma_5)d_\alpha )
\sum_q e_q (\bar{q}_\beta \gamma^\mu (1-\gamma_5)q_\beta ) \\ 
O_{10} & = & \frac{3}{2}(\bar{s}_\alpha \gamma_\mu (1-\gamma_5)d_\beta )
\sum_q e_q (\bar{q}_\beta \gamma^\mu (1-\gamma_5)q_\alpha ) 
\label{eq:ops10}
\end{eqnarray}
Here $\alpha$ and $\beta$ are color indices, $e_q$ is quark
electric charge, and summation is done over all light quarks. 

Isospin amplitudes are defined as 
\begin{equation}
\label{amp}
A_{0,2}e^{i\delta_{0,2}} \equiv \langle (\pi\pi )_{I=0,2}|H_W|K^0\rangle ,
\end{equation}
where $\delta_{0,2}$ are the final state interaction phases of the
two channels. Experimentally
\begin{equation}
\omega = \Re A_0 /\Re A_2 \simeq 22 \, .
\end{equation}

Direct CP violation parameter $\varepsilon '$ is defined in terms 
of imaginary parts of these amplitudes:
\begin{equation}
\varepsilon ' = -\frac{\Im A_0 - \omega \Im A_2}{\sqrt{2}\omega\Re A_0}
 e^{i(\pi/2 + \delta_2 - \delta_0)}.
\end{equation}
Experiments are measuring the quantity $\Re \varepsilon '/\varepsilon$, 
which is given by
\begin{equation}
\label{eq:epsp}
\Re \,\frac{\varepsilon '}{\varepsilon} \simeq
\frac{G_F}{2\omega |\varepsilon |\Re{A_0}} \,
\mbox{Im}\, \lambda_t \, \,
 \left[ \Pi_0 - \omega \: \Pi_2 \right] ,
\end{equation}
where
\begin{eqnarray}
\label{P0}
 \Pi_0 & = &  \sum_i y_i \, \langle (\pi\pi )_{I=0}|O_i^{(0)}|K^0\rangle 
(1 - \Omega_{\eta +\eta '}) \\
\label{P2}
 \Pi_2 & = &  \sum_i y_i \, \langle (\pi\pi )_{I=2}|O_i^{(2)}|K^0\rangle  
\end{eqnarray}
with $\mbox{Im}\, \lambda_t \equiv \Im V_{td}V^*_{ts}$, and where
$\Omega_{\eta + \eta'} \sim 0.25\pm 0.05$ takes into account the effect
of isospin breaking in quark masses ($m_u \neq m_d$). $O_i^{(0)}$ and
$O_i^{(2)}$ are isospin 0 and 2 parts of the basis operators.
Their expressions are given in the Appendix for completeness.

\subsection{Treatment of charm quark}

The effective Hamiltonian given above is obtained in the continuum 
theory in which the top, bottom and charm quarks are integrated out. 
(In particular, the summation in Eqs.~(\ref{eq:ops3}--\ref{eq:ops10}) 
is done over $u$, $d$ and $s$ quarks.) This makes sense only when
the scale $\mu$ is sufficiently low compared to the charm quark mass.
As mentioned in Ref.~\cite{charm}, at scales comparable to $m_c$ 
higher-dimensional 
operators can contribute considerably. Then one should consider
an expanded set of operators including those containing the charm quark.
Lattice treatment of the charm quark is possible but
in practice quite limited, for example by having to work at much
smaller lattice spacings and having a more complicated set
of operators and contractions. Therefore
we have opted to work in the effective theory in which the charm quark
is integrated out. Since we typically use $\mu \sim 2$~GeV in our
simulations, this falls into a dangerous region. We hope that
the effects of higher-dimensional operators can still be neglected, but
strictly speaking this issue should be separately investigated.

\subsection{Calculating $\langle \pi\pi|O_i|K^0\rangle$.}

As was shown by Martinelli and Testa~\cite{testa}, two-particle
hadronic
states are very difficult to construct on the lattice (and in general,
in any Euclidean description). We have
to use an alternative procedure to calculate the matrix elements 
appearing in Eqs.~(\ref{amp},\ref{P0},\ref{P2}).
We choose the method ~\cite{bernard} in which the lowest-order
chiral perturbation theory is used to relate 
$\langle \pi\pi |O_i|K^0\rangle$ to matrix elements involving one-particle states:
\begin{eqnarray}
\label{eq:chpt1}
\langle \pi^+\pi^-|O_i|K^0\rangle & = & \frac{m_K^2-m_\pi^2}{f}\gamma \\
\langle \pi^+|O_i|K^+\rangle & = & (p_\pi \cdot p_K)\gamma - 
                \frac{m_s+m_d}{f}\delta \\
\label{eq:chpt3}       
\langle 0|O_i|K^0\rangle & = & (m_s-m_d)\delta  ,
\end{eqnarray}
where $f$ is the lowest-order pseudoscalar decay constant.
The masses in the first of these formulae are the physical meson masses,
while the quark masses and the momenta in the second and third formulae
are meant to be from actual simulations on the lattice
(done with unphysical masses). 
These relationships ignore higher order terms in the chiral expansion, 
most importantly the final state interactions. 
Therefore this method suffers from a significant uncertainty. 
Golterman and Leung~\cite{golterman} have computed one-loop correction 
for $\Delta I=3/2$ amplitude in chiral perturbation theory. 
They find this correction can be large, up to 30\% or 60\%, depending
on the values of unknown contact terms and the cut-off.   

\section{Lattice techniques}
\label{sec:lattice details}

\subsection{Mixing with lower-dimensional operators.}

Eqs.~(\ref{eq:chpt1}--\ref{eq:chpt3}) handle unphysical
$s \leftrightarrow d$ mixing in $\langle\pi^+|O_i|K^+\rangle$ 
by subtracting the unphysical part proportional to 
$\langle 0|O_i|K^0\rangle$. This is equivalent to subtracting
the operator 
\begin{equation}
O_{sub} \equiv (m_d+m_s)\bar{s}d + (m_d-m_s)\bar{s}\gamma_5d \,.
\label{eq:SubOp}
\end{equation}
As shown by Kilcup, Sharpe {\it et al.} in Refs.~\cite{ToolKit,WeakME}, 
these statements are also true
on the lattice if one uses staggered fermions. A number of Ward identities
discussed in these references show that lattice formulation with
staggered fermions retains 
the essential chiral properties of the continuum theory. In particular,
$O_{sub}$ defined in Eq.~\ref{eq:SubOp} is the only lower-dimensional 
operator appears in mixing with the basis operators.
(Lower-dimensional operators have to be subtracted non-perturbatively
since they are multiplied by powers of $a^{-1}$.)
We employ the non-perturbative procedure suggested in Ref.~\cite{WeakME}:
\begin{equation}
\label{eq:sub1}
\langle \pi^+\pi^-|O_i|K^0\rangle = 
\langle \pi^+|O_i - \alpha_i O_{sub}|K^+\rangle \cdot \frac{m_K^2-m_\pi^2}
{(p_\pi\cdot p_K)f} \, , 
\end{equation}
where $\alpha_i$ are found from 
\begin{equation}
0 = \langle 0|O_i - \alpha_i O_{sub}|K^0\rangle \, .
\label{eq:sub2}
\end{equation}
This procedure is equivalent to the lattice version of 
Eqs.~(\ref{eq:chpt1}--\ref{eq:chpt3}) and allows subtraction 
timeslice by timeslice.

Throughout our simulation we use only degenerate mesons, i.e. $m_s=m_d=m_u$.
Since only negative parity part of $O_{sub}$ contributes in 
Eq.~(\ref{eq:sub2}), one naively expects infinity when calculating 
$\alpha_i$. However, matrix elements 
$\langle 0|O_i|K^0\rangle$ of all basis operators 
vanish when $m_s=m_d$ due to invariance of both the Lagrangian
and all the operators in question under the CPS symmetry, which
is defined as the CP symmetry combined with interchange of $s$ and $d$ 
quarks. Thus calculation of $\alpha_i$ requires taking the first derivative 
of $\langle 0|O_i|K^0\rangle$ with respect to $(m_d-m_s)$. In order
to evaluate the first derivative numerically, we insert another
fermion matrix inversion in turn into all propagators involving
the strange quark. Detailed expressions for all contractions 
are given in the Appendix.

\begin{figure}[p]
\begin{center}
\leavevmode
\centerline{\epsfysize=12cm \epsfbox{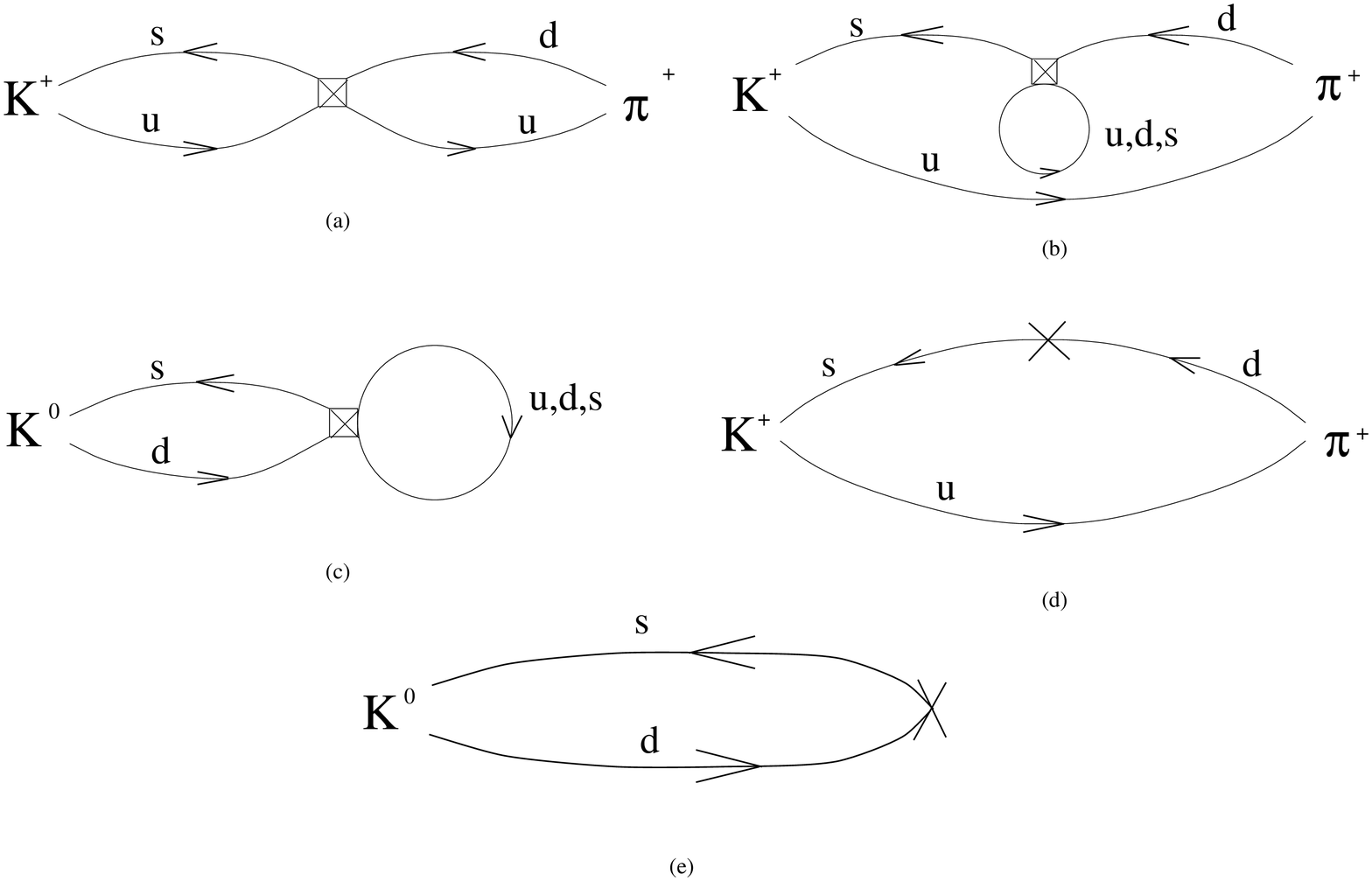}}
\end{center}
\caption{Five diagrams types needed to be computed: (a) ``Eight'';
(b) ``Eye''; (c) ``Annihilation''; (d) ``Subtraction''; (e)
two-point function.}
\label{diagrams}
\end{figure} 

\subsection{Diagrams to be computed}

\label{sec:diag}

According to Eqs.~(\ref{eq:sub1},\ref{eq:sub2}), we need to compute three
diagrams involving four-fermion operators (shown in Fig.~\ref{diagrams})
and a couple of bilinear contractions. The ``eight'' contraction type 
(Fig.~\ref{diagrams}a) is relatively cheap to compute. It is the only
contraction needed for the $\Delta I=3/2$ amplitude.
The ``eye'' and ``annihilation'' 
diagrams (Fig.~\ref{diagrams}b and~\ref{diagrams}c) are much more 
expensive since they involve calculation of propagators from
every point in space-time.

\subsection{Lattice parameters and other details}

The parameters of simulation are listed in the Table~\ref{tab:parameters}.
We use periodic boundary conditions in both space and time.
Our main ``reference'' ensemble is a set of quenched configurations
at $\beta \equiv 6/g^2 =6.0$ ($Q_1$). In addition, we use an
ensemble with a larger lattice volume ($Q_2$), an ensemble
with $\beta =6.2$ ($Q_3$) for checking the lattice spacing dependence,
and an  ensemble with 2 dynamical flavors ($m=0.01$) generated by the 
Columbia group, used for checking the impact of quenching. 
The ensembles were obtained using 4 sweeps of $SU(2)$ overrelaxed
and 1 sweep of $SU(2)$ heatbath algorithm\footnote{except for the dynamical 
set which was obtained by R-algorithm~\cite{Columbia1}}. The configurations 
were separated by
1000 sweeps, where one sweep includes three $SU(2)$ subgroups updates.

\begin{table}[tbh]
\caption{Simulation parameters}
\label{tab:parameters}
\begin{tabular}{ccccccc}
\hline\hline
Ensemble & $N_f$ & $\beta $ & Size & L, fm & Number of & Quark masses\\
name & & & & & configurations & used \\
\tableline
$Q_1$ & 0 & 6.0 & $16^3\times (32\times 4)$ & 1.6 & 216 & 0.01 --- 0.05 \\
$Q_2$ & 0 & 6.0 & $32^3\times (64\times 2)$ & 3.2 & 26 & 0.01 --- 0.05 \\
$Q_3$ & 0 & 6.2 & $24^3\times (48\times 4)$ & 1.7 & 26 & 0.005 --- 0.03 \\
$D$   & 2 & 5.7 & $16^3\times (32\times 4)$ & 1.6 & 83 & 0.01 --- 0.05 \\
\hline\hline
\end{tabular}
\end{table}

We use the standard staggered fermion action. 
Fermion matrices are inverted by conjugate gradient.
Jackknife is used for statistical analysis. 

As explained below, we have extended the lattice 4
times\footnote{for all ensemble except the biggest volume, 
which we extend 2 times.}
in time dimension by copying gauge links. This is done in order to 
get rid of excited states contamination and wrap-around effects. 

The lattice spacing values for quenched ensembles were obtained by 
performing a fit in
the form of the asymptotic scaling to the quenched data of $\rho$ meson
mass given elsewhere~\cite{spectrum}. Lattice spacing for the dynamical 
ensemble is also set by the $\rho$ mass~\cite{Columbia}. 

Some other technicalities are as follows.
We work in the two flavor formalism. We use local wall sources
that create pseudoscalar mesons at rest.
(Smearing did not have a substantial effect.) 
The mesons are degenerate ($m_s=m_d=m_u$, $m_\pi=m_K$).
We use staggered
fermions and work with gauge-invariant operators, since the 
gauge symmetry enables significant reduction of the list of 
possible mixing operators. The staggered flavour structure
is assigned depending on the contraction type.
Our operators are tadpole-improved. This 
serves to `improve'' the perturbative expansion at a later stage 
when we match the lattice and continuum operators.
For calculating fermion loops we employ the $U(1)$ pseudofermion
stochastic estimator. 
More details and explanation of some of these 
terms can be found in the Appendix.

\begin{figure}[!hbt]
\begin{center}
\leavevmode
\centerline{\epsfysize=7cm \epsfbox{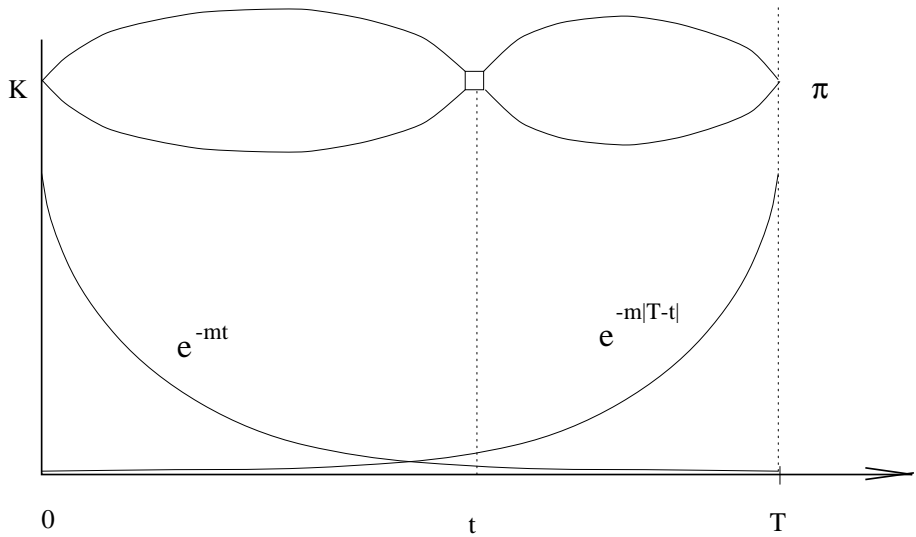}}
\end{center}
\caption{The general setup of calculation of
$\langle \pi^+|O_i|K^+\rangle$ (without loss of generality, 
an ``eight'' contraction is shown). The kaon source is
at the timeslice 0, while the pion sink is at the timeslice $T$.
The operator is inserted at a variable time $t$. The result of this
contraction is proportional to the product of two exponentials
shown in the figure.}
\label{setup}
\end{figure} 

\subsection{Setup for calculating matrix elements of four-fermion \\
operators}

Consider the setup for calculation of $\langle \pi^+|O_i|K^+\rangle$.
Kaons are created at $t_0=0$, the operators are inserted at 
a variable time $t$, and the pion sink is located at the
time $T$ (see Fig.~\ref{setup}), where $T$ is sufficiently large.
In principle, a number of states with pseudoscalar quantum numbers
can be created by the kaon source.
Each state's contribution is proportional to $\sqrt{Z}e^{-m|t|}$, so the 
lightest state (kaon) dominates at large enough $t$.
Analogously, states annihilated by the sink contribute proportionally
to $\sqrt{Z}e^{-m|T-t|}$, which is dominated by the pion.

In this work kaon and pion have equal mass.
In the middle of the lattice, where $t$
is far enough from both 0 and $T$, we expect to see a plateau, 
corresponding to $Z e^{-m_\pi T}\langle\pi|O|K\rangle$. 
This plateau is our working region (see Fig.~\ref{plateau}). 

\begin{figure}[!htb]
\begin{center}
\leavevmode
\centerline{\epsfysize=6.5cm \epsfbox{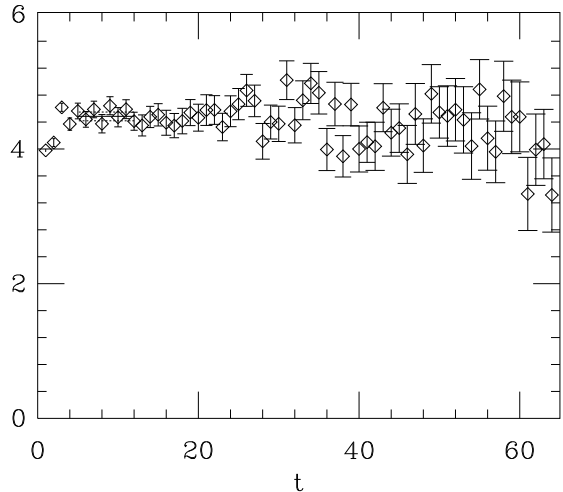}}
\end{center}
\caption{An example of the signal for 
$\langle 0|O_2|K^0\rangle\,/\,
[(m_d-m_s)\,\langle 0|\overline{s}\gamma_5 d|K^0\rangle]$ on $Q_1$
ensemble. The kaon source is at $t=1$. We average over the range
of $t$ from 5 to 12 (shown with horizontal lines).}
\label{ann}
\end{figure}

As concerns the kaon annihilation matrix elements 
$\langle 0|O_i|K^0\rangle$, we only need their ratio to 
$\langle 0|\overline{s}\gamma_5 d|K^0\rangle$, in which the 
factors $\sqrt{Z}e^{-mt}$ cancel. Indeed, we observe a rather steady
plateau (Fig.~\ref{ann}).  

\begin{figure}[!bht]
\begin{center}
\leavevmode
\centerline{\epsfysize=5cm \epsfbox{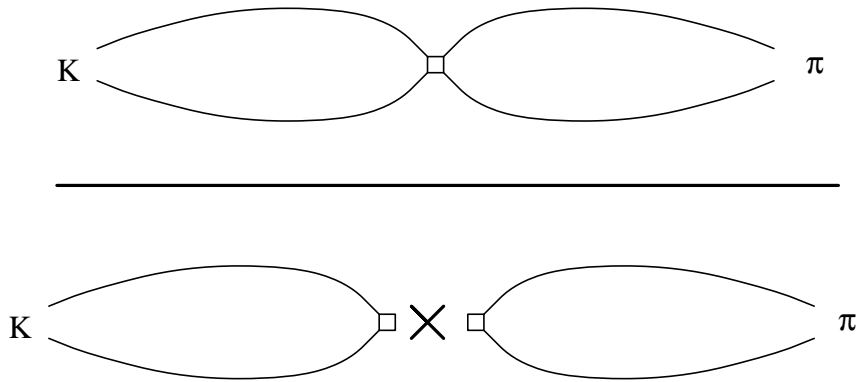}}
\end{center}
\caption{$B$ ratio is formed by dividing the four-fermion matrix element
by the product of two-point functions, typically involving $A_\mu$
or $P$ bilinears. All the operators involved are inserted at the same
timeslice $t$, and summation is done over spatial volume. The external 
meson sources 
are located at timeslices 0 and $T$, both in the numerator and the 
denominator. This enables cancellation of some common factors.}
\label{ratio}
\end{figure} 

\subsection{$B$ ratios}

It has become conventional
to express the results for matrix elements
in terms of so-called $B$ ratios, which are the ratios of desired 
four-fermion matrix elements to their
values obtained by vacuum saturation approximation (VSA).
For example, the $B$ ratios of operators $O_2$ and $O_4$ are formed by
dividing the full matrix element by the product of axial-current
two-point functions (Fig.~\ref{ratio}).
We expect the denominator to form a plateau 
in the middle of the lattice, equal to
$Z e^{-m_\pi T} \, \langle\pi|A_\mu|0\rangle \,\cdot \,
\langle 0|A^\mu|K\rangle$,
where $A^\mu$ are the axial vector currents with appropriate flavor quantum
numbers for kaon and pion. The
factor $Z e^{-m_\pi T}$ cancels, leaving the desirable ratio
$\langle\pi|O|K\rangle \, / \,
(\langle\pi|A_\mu|0\rangle\, \cdot \, \langle 0|A^\mu|K\rangle)$. 
Apart from common normalization factors, 
a number of systematic uncertainties also tend to cancel in this ratio,
including the uncertainty in the lattice spacing, quenching and
in some cases perturbative correction uncertainty. 
Therefore, it is sometimes reasonable to give lattice answers in terms
of the $B$ ratios. 

However, eventually the physical matrix element
needs to be reconstructed by using the known experimental parameters 
(namely $f_K$) to compute VSA. In some cases, such as for operators
$O_5$---$O_8$, the VSA itself is known very imprecisely due to the
failure of perturbative matching (see Sec.~\ref{sec:pert}).
Then it is more reasonable to give answers in terms of matrix elements
in physical units. We have adopted the strategy of expressing all matrix 
elements in units of $\langle\pi|A_\mu|0\rangle \, \langle 0|A^\mu|K\rangle
= (f_K^{latt})^2 m_M^2$ at an intermediate stage, and using 
pre-computed $f_K^{latt}$ at the given meson mass to convert to physical 
units. This method is sensitive to the choice of the lattice spacing. 

\begin{figure}[!hbt]
\begin{center}
\leavevmode
\centerline{\epsfysize=7cm \epsfbox{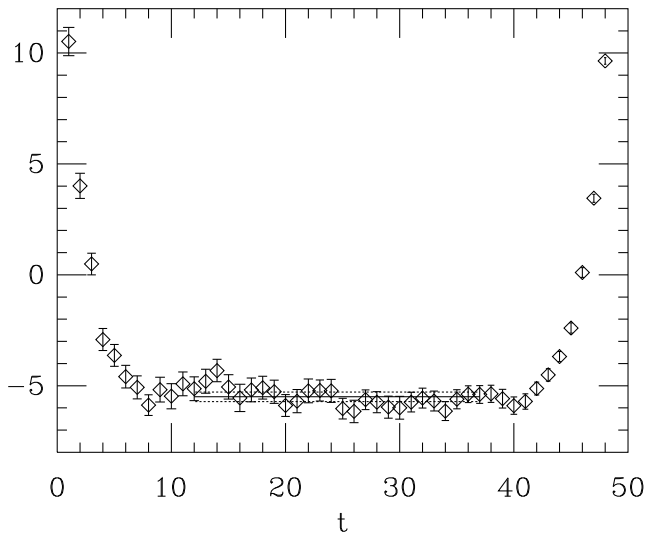}}
\end{center}
\caption{An example of the signal we get for one of the $B$ ratios
(in this case, for the ``eye'' part of the $O_2$ operator on $Q_1$ ensemble). 
The wall sources are at $t=1$ and $t=49$. We see that
the excited states quickly disappear and a stable, well-distinguished
plateau is observed. We perform jackknife averaging in the range of $t$
from 12 to 37 (shown with the horizontal lines). It is important to 
confirm the existence of the plateau for reliability of the results.}
\label{plateau}
\end{figure} 

It is very important to check that the time distance between the
kaon and pion sources $T$ is large enough so 
that the excited states do not contribute. That is, the plateau
in the middle of the lattice should be sufficiently flat,
and the $B$ ratios should not depend on $T$. We have found that 
in order to satisfy this requirement the lattice has to be
artificially extended in time direction by using
a number of copies of the gauge links (4 in the case of the small
volume lattices, 2 otherwise). We are using $T=72$ for $Q_3$ 
($\beta =6.2$) ensemble, and $T=48$ for the rest.
An example of a plateau that we obtain
with this choice of $T$ is shown in Fig.~\ref{plateau}.
To read off the result, we average over the whole extension
of the plateau, and use jackknife to estimate the statistical
error in this average.

\section{$\Delta I=1/2$ rule results}
\label{sec:di12}

Using the data obtained for matrix elements of basis operators,
in this section we report numerical results for $\Re A_0$
and $\Re A_2$ amplitudes as well as their ratio. We discuss these
amplitudes separately since the statistics for $\Re A_2$ is
much better and the continuum limit extrapolation is possible. 

\subsection{$\Re A_2$ results}
\label{sec:A2}

The expression for $\Re A_2$ can be written as
\begin{equation}
\Re A_2 = \frac{G_F}{\sqrt{2}}\, V_{ud}V_{us}^*\, z_+(\mu )
\langle O_2\rangle _2,
\end{equation}
where $z_+ (\mu )$ is a Wilson coefficient and
\begin{equation}
\langle O_2\rangle _2 \equiv \langle (\pi\pi)_{I=2}|O_2^{(2)}|K\rangle .
\end{equation}
Here
\begin{eqnarray}
O_2^{(2)} = O_1^{(2)} & = & \frac{1}{3} 
[ (\overline{s}\gamma_\mu(1-\gamma_5)u)(\overline{u}\gamma^\mu(1-\gamma_5)d) 
\nonumber \\ & &
+(\overline{s}\gamma_\mu(1-\gamma_5)d)(\overline{u}\gamma^\mu(1-\gamma_5)u)
-(\overline{s}\gamma_\mu(1-\gamma_5)d)(\overline{d}\gamma^\mu(1-\gamma_5)d).
\end{eqnarray}
In the lowest order chiral perturbation theory the matrix element
$\langle O_2\rangle_2$ can be expressed as
\begin{equation}
\langle O_2\rangle _2 
= \sqrt{2} \,\frac{m_K^2-m_\pi^2}{f} 
\,\frac{\langle\pi^+|O_2^{(2)}|K^+\rangle}{m^2}.
\end{equation}
The latter matrix element involves only ``eight'' diagrams. Moreover, 
in the limit of preserved $SU(3)_{\mathrm flavor}$ symmetry
it is directly related~\cite{donoghue} to parameter $B_K$ (which is the 
$B$ ratio of the neutral kaon mixing operator  
$O_K= (\overline{s}\gamma_L d) \;(\overline{s}\gamma_L d)$), so that
\begin{equation}
\langle O_2\rangle _2 
= \frac{4\sqrt{2}}{9} \,\frac{m_K^2-m_\pi^2}{f_{\mbox{exp}}} 
\,f_{\mbox{latt}}^2\,B_K \, ,
\end{equation}

\begin{figure}[htb]
\begin{center}
\leavevmode
\centerline{\epsfysize=6cm \epsfbox{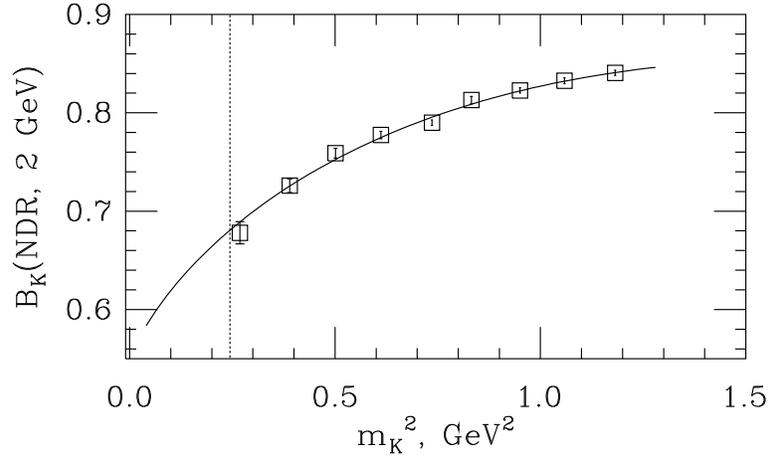}}
\end{center}
\caption{Parameter $B_K$ in NDR $\overline{\mathrm MS}$ scheme at 2 GeV
on the dynamical ensemble vs. the meson mass squared. The fit
is of the form \mbox{$B_K=a+bm_K^2+c\;m_K^2\log{m_K^2}$.} The vertical line
here and in the other plots below marks the physical kaon mass.}
\label{Bk}
\end{figure}

\begin{figure}[tbh]
\begin{center}
\leavevmode
\centerline{\epsfysize=6cm \epsfbox{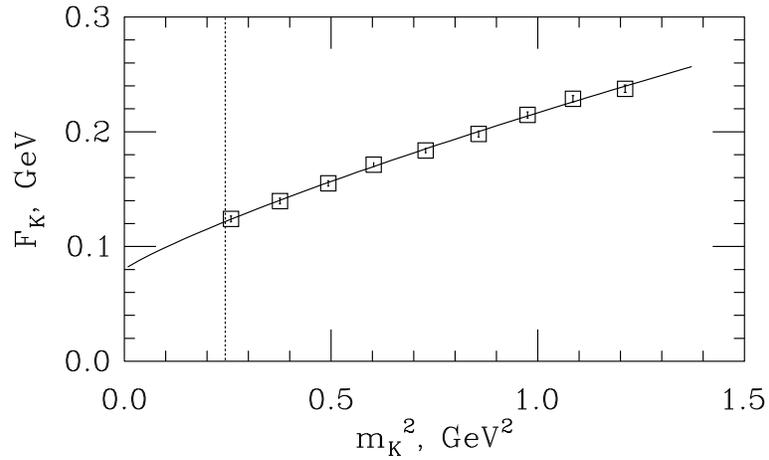}}
\end{center}
\caption{Pseudoscalar decay constant ($F_\pi = 93$ MeV experimentally) on 
the dynamical ensemble vs. meson mass squared. The fit is of the same 
form as $B_K$.}
\label{fk}
\end{figure}

\begin{figure}[htb]
\begin{center}
\leavevmode
\centerline{\epsfysize=8cm \epsfbox{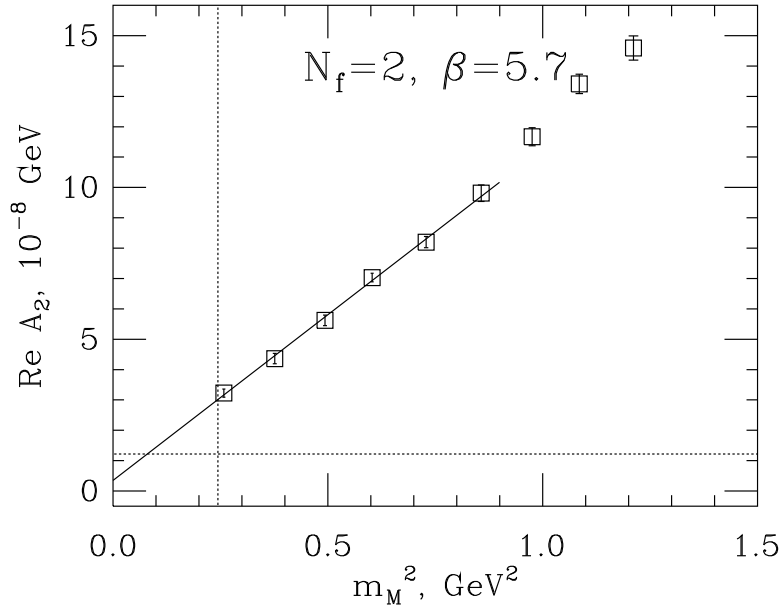}}
\end{center}
\caption{$\Re A_2$ for the dynamical ensemble. The fit is of
the same form as $B_K$. The horizontal line is the experimental value
of 1.23 GeV. }
\label{A2}
\end{figure}

Parameter $B_K$ is rather well studied (for example, by 
Kilcup, Pekurovsky~\cite{PK1} and JLQCD collaboration~\cite{jlqcd}).
Quenched chiral perturbation theory~\cite{sharpe1} predicts 
the chiral behaviour of the form 
\mbox{$B_K=a+bm_K^2+c\;m_K^2\log{m_K^2}$,} which 
fits the data well (see Fig.~\ref{Bk}) and
yields a finite non-zero value in the chiral limit. 
Note that $\Re A_2$ is proportional
to the combination $B_K f_{\mbox{latt}}^2$, and since both multipliers 
have a significant
dependence on the meson mass (Figs.~\ref{Bk} and~\ref{fk}), 
$\Re A_2$ is very sensitive to that mass.
Fig.~\ref{A2} shows $\Re A_2$ data for the dynamical ensemble, based on
$B_K$ values we have reported elsewhere~\cite{PK1}. 
Which meson mass
should be used to read off the result becomes an open question. 
If known, the higher order chiral terms would remove this ambiguity.
Forced to make a choice, we 
extrapolate to \mbox{$M^2=(m_K^2+m_\pi^2)/2$}.
Using our data for $B_K$ in quenched QCD and taking
the continuum limit we obtain:
$\Re A_2 = (1.7 \pm 0.1)\cdot 10^{-8}\;\mbox{GeV}$,
where the error is only statistical,
to be compared with the experimental result
\mbox{$\;\Re A_2 = 1.23 \cdot 10^{-8}\;\mbox{GeV}$. }

\clearpage

Higher order chiral terms (including the meson mass dependence)
are the largest systematic error in this determination.
According to Golterman and Leung~\cite{golterman}, one-loop corrections
in (quenched) chiral perturbation theory are expected to be
as large as $30\%$ or $60\%$. Other uncertainties (from lattice spacing 
determination, from perturbative operator matching and
from using finite lattice volume) are much smaller.


\subsection{$\Re A_0$ results}

Using Eqs.~(\ref{eq:sub1},\ref{eq:sub2}), $\Re A_0$ can be expressed 
as\footnote{In our normalization $\Re A_0 = 27.2 \cdot 10^{-8}$.}
\begin{equation}
\Re A_0 = \frac{G_F}{\sqrt{2}}V_{ud}V_{us}^* \frac{m_K^2-m_\pi^2}{f}
\sum_i z_i R_i ,
\end{equation}
where $z_i$ are Wilson coefficients and 
$$
R_i \equiv \frac{\langle \pi^+|O_i^{(0)}|K^+\rangle_s}{m^2}.
$$
The subscript '$s$' indicates that these matrix elements already
include subtraction of \linebreak $\langle \pi^+|O_{sub}|K^+\rangle$. 
All contraction types are needed, including the expensive ``eyes''
and ``annihilations''.
$O_i^{(0)}$ are isospin 0 parts of operators 
$O_i$ (given in the Appendix for completeness). For example,
\begin{eqnarray}
O_1^{(0)} & = & \frac{2}{3} 
(\overline{s}\gamma_\mu (1-\gamma_5)d)(\overline{u}\gamma^\mu (1-\gamma_5)u)
-\frac{1}{3}(\overline{s}\gamma_\mu (1-\gamma_5)u)(\overline{u}\gamma^\mu 
(1-\gamma_5)d)  \nonumber \\
& + & \frac{1}{3}(\overline{s}\gamma_\mu (1-\gamma_5)d)
(\overline{d}\gamma^\mu (1-\gamma_5)d) \\
O_2^{(0)} & = & \frac{2}{3} 
(\overline{s}\gamma_\mu (1-\gamma_5)u)(\overline{u}\gamma^\mu (1-\gamma_5)d)
-\frac{1}{3}(\overline{s}\gamma_\mu (1-\gamma_5)d)(\overline{u}\gamma^\mu 
(1-\gamma_5)u) \nonumber \\
& + & \frac{1}{3}(\overline{s}\gamma_\mu (1-\gamma_5)d)
(\overline{d}\gamma^\mu (1-\gamma_5)d) 
\end{eqnarray}

\begin{figure}[htbp]
\begin{center}
\leavevmode
\centerline{\epsfysize=10cm \epsfbox{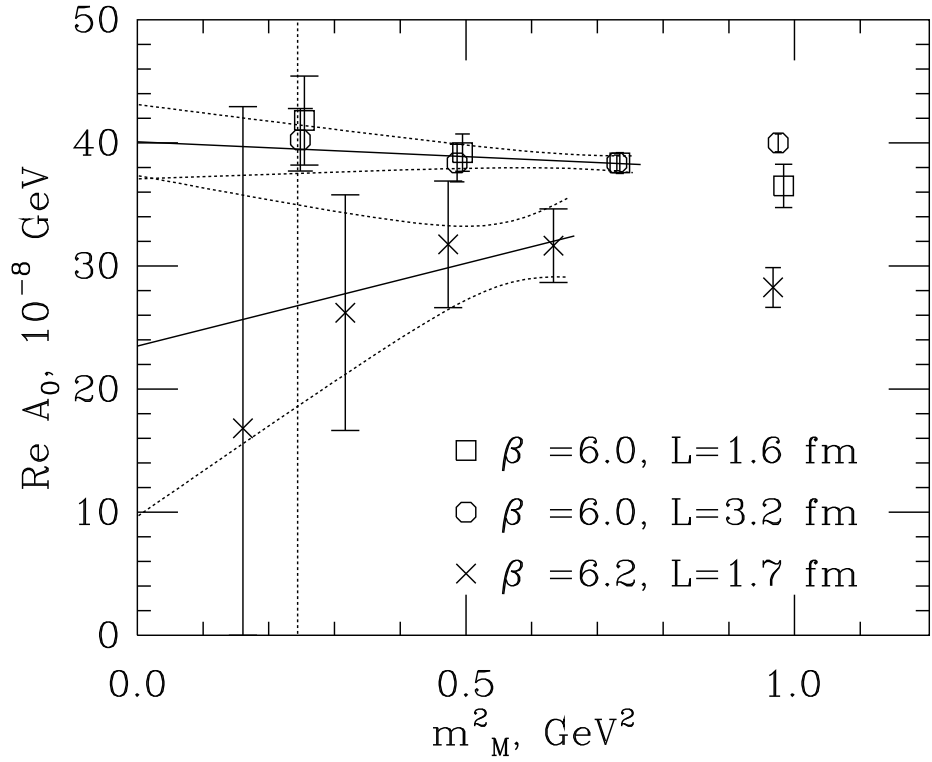}}
\end{center}
\caption{$\Re A_0$ for quenched ensembles plotted against the meson mass 
squared. The upper group of points
is for ensembles $Q_1$ and $Q_2$, while the lower group is for $Q_3$. 
Only statistical errors are shown. }
\label{A0}
\end{figure}

\begin{figure}[tbh]
\begin{center}
\leavevmode
\centerline{\epsfysize=6cm \epsfbox{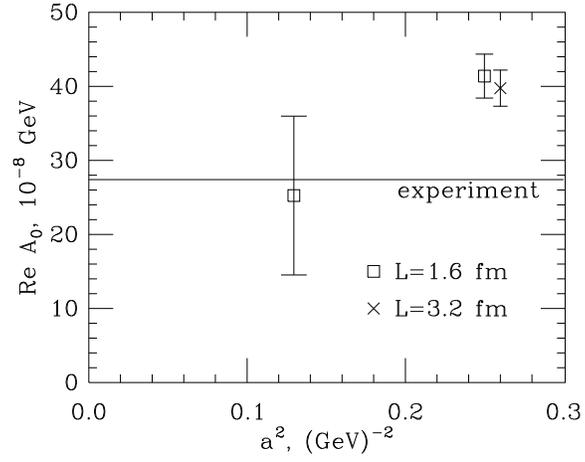}}
\end{center}
\caption{$\Re A_0$ for quenched ensembles plotted against lattice spacing
squared. The horizontal line shows the experimental result of 
$27.2\cdot 10^{-8}$ GeV. Only statistical errors are shown.}
\label{A0cont}
\end{figure}

The results for quenched $\beta =6.0$ and $\beta =6.2$ ensembles
are shown in Fig.~\ref{A0}. Dependence on the 
meson mass is small, so there is no big ambiguity about the mass
prescription as in the $\Re A_2$ case. 
Some lattice spacing dependence may be present (Fig.~\ref{A0cont}), 
although the statistics
for $\beta =6.2$ ensemble is too low at this moment.

The effect of
the final state interactions (contained in the higher order
terms of the chiral perturbation theory) is likely to be large.
This is the biggest and most poorly estimated
uncertainty. 

An operator matching uncertainty arises due to mixing of $O_2$ with $O_6$ 
operator through penguin diagrams in lattice perturbation
theory. This is explained in the Section~\ref{sec:A0pert}. We estimate
this uncertainty at 20\% for all ensembles.  

\begin{figure}[htb]
\begin{center}
\leavevmode
\centerline{\epsfysize=6cm \epsfbox{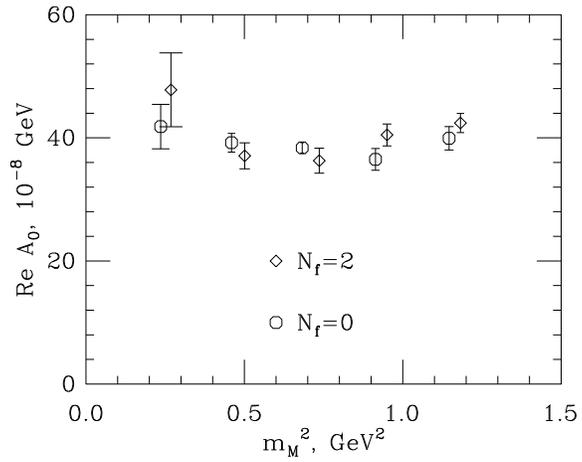}}
\end{center}
\caption{Comparison of quenched ($Q_1$) and dynamical results for $\Re A_0$
at comparable lattice spacings.}
\label{A0quench}
\end{figure}

As for other uncertainties, we have checked the lattice volume
dependence by comparing ensembles $Q_1$ and $Q_2$ (1.6 and 3.2 fm
at $\beta =6.0$).
The dependence was found to be small, so we consider $(1.6 \;{\mathrm fm})^3$ 
as a volume large enough to hold the system. We have also checked the effect
of quenching and found it to be small compared to noise
(see Fig.~\ref{A0quench}). 

\begin{figure}[!tbh]
\begin{center}
\leavevmode
\centerline{\epsfysize=6cm \epsfbox{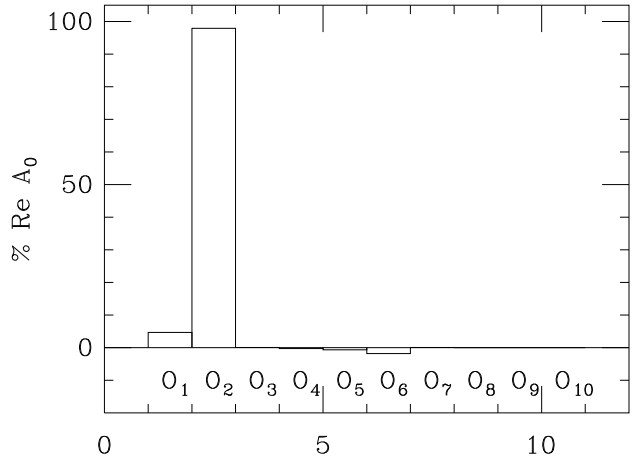}}
\end{center}
\caption{Contribution of various operators to $\Re A_0$.}
\label{A0hist}
\end{figure}

The breakdown of contributions of various basis
operators to $\Re A_0$ is shown in Fig.~\ref{A0hist}.
By far, $O_2$ plays the most important role, whereas penguins
have only a small influence. 

\clearpage

\begin{figure}[!bth]
\begin{center}
\leavevmode
\centerline{\epsfysize=10cm \epsfbox{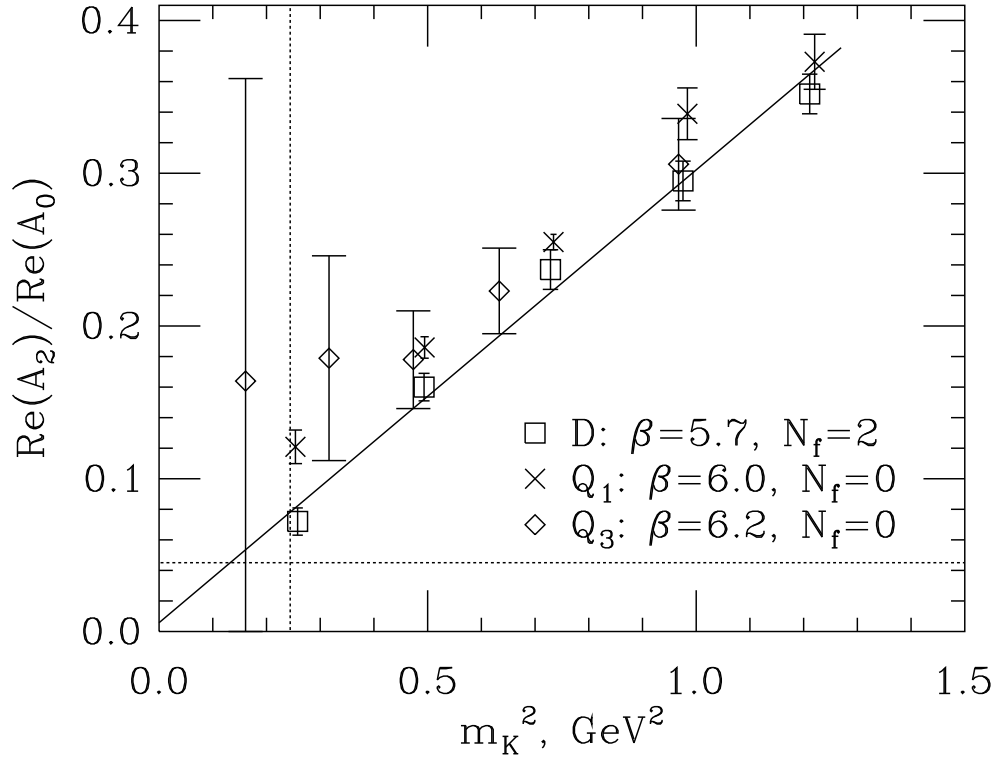}}
\end{center}
\caption{The ratio $\Re A_2/\Re A_0$ versus the meson mass squared
for quenched and dynamical ensembles.
Ensembles $Q_1$ and $D$ have comparable lattice spacings. 
The dynamical ensemble data were used for the fit. 
The big slope of the fit line is accounted for by the
mass dependence of $\Re A_2$. The horizontal line shows the experimental
value of $1/22$. The error bars show only the statistical errors 
obtained by jackknife.}
\label{omega}
\end{figure} 

\subsection{Amplitude ratio}

Shown in Fig.~\ref{omega} is the ratio $\Re A_2/\Re A_0$ as directly 
computed on the lattice for quenched and dynamical data sets.
The data exhibit strong dependence on
the meson mass, due to $\Re A_2$ chiral behaviour (compare with 
Fig.~\ref{A2}). 

Within our errors the results seem to confirm, indeed, the 
common belief that most of the $\Delta I=1/2$ enhancement comes 
from the ``eye'' and ``annihilation'' diagrams. The exact amount
of enhancement is broadly consistent with experiment while being
subject to considerable uncertainty due to higher-order chiral terms.
Other systematic errors are the same as those described 
in the previous Subsection.

\section{Operator matching}

\label{sec:pert}

As mentioned before, we have computed the matrix elements of all 
relevant operators with an acceptable statistical accuracy.
These are regulated in the lattice renormalization 
scheme. To get physical results, 
operators need to be matched to the same scheme in which the Wilson
coefficients were computed in the continuum, namely $\overline{\mathrm MS}$
NDR. While perturbative matching works quite well for
$\Re A_0$ and $\Re A_2$, it seems to break down severely for
matching operators relevant for $\varepsilon '/\varepsilon$.

\subsection{Perturbative matching and $\Re A_0$}

\label{sec:A0pert}

Conventionally, lattice and continuum operators are matched using
lattice perturbation theory:
\begin{equation}
\displaystyle
\label{eq:matching}
O_i^{\it cont}(q^*) =  O_i^{\it lat} + \displaystyle\frac{g^2(q^*a)}{16\pi^2}\displaystyle\sum_j(\gamma_{ij}\ln (q^*a)
 + C_{ij})O_j^{\it lat} + O(g^4) + O(a^n) ,
\end{equation}
where $\gamma_{ij}$ is the one-loop anomalous dimension matrix 
(the same in the continuum 
and on the lattice), and $C_{ij}$ are finite coefficients calculated
in one-loop lattice perturbation theory~\cite{Ishizuka,PatelSharpe}. 
We use the ``horizontal matching'' 
procedure~\cite{horizontal}, whereby the same coupling constant
as in the continuum ($g_{\overline{MS}}$) is used.
The operators are matched at an intermediate scale 
$q^*$ and evolved using the continuum renormalization
group equations to the reference scale $\mu$, which we take 
to be 2 GeV.

In calculation of $\Re A_0$ and $\Re A_2$, the main contributions
come from left-left operators. One-loop renormalization
factors for such (gauge-invariant) operators were computed by
Ishizuka and Shizawa~\cite{Ishizuka} (for current-current diagrams)
and by Patel and Sharpe~\cite{PatelSharpe} (for penguins). 
These factors are fairly small, so at the first glance the perturbation theory
seems to work well, in contrast to the case of left-right operators 
essential for estimating $\varepsilon '/\varepsilon$, as described below.
However, even in the case of $\Re A_0$ there is a certain
ambiguity due to mixing of $O_2$ operator with $O_6$ through
penguin diagrams. The matrix element of $O_6$ is rather large, so
it heavily influences $\langle O_2\rangle$ in spite of the small
mixing coefficient. Operator $O_6$ receives enormous renormalization 
corrections in the first order, as discussed below. Therefore, there
is an ambiguity as to whether the mixing should be evaluated
with renormalized or bare $O_6$. Equivalently, the higher-order
diagrams (such as Fig.~\ref{higher-order}b and~\ref{higher-order}d) 
may be quite important.  

In order to estimate the uncertainty of neglecting higher-order diagrams,
we evaluate the mixing with $O_6$ renormalized
by the partially non-perturbative procedure described below, and
compare with results obtained by evaluating mixing with bare $O_6$.
The first method amounts to resummation of those higher-order diagrams 
belonging to type (b) in Fig.~\ref{higher-order}, while the second method
ignores all higher-than-one-order corrections. 
Results quoted in the previous Section
were obtained by the first method, which is also close to using
tree-level matching. The second method would produce
values of $\Re A_0$ lower by about 20\%.
Thus we consider 20\% a conservative estimate of the matching uncertainty.  

In calculating $\varepsilon '/\varepsilon$ the operator 
matching issue becomes a much more serious obstacle as explained below.

\begin{figure}[htb]
\begin{center}
\leavevmode
\centerline{\epsfysize=8cm \epsfbox{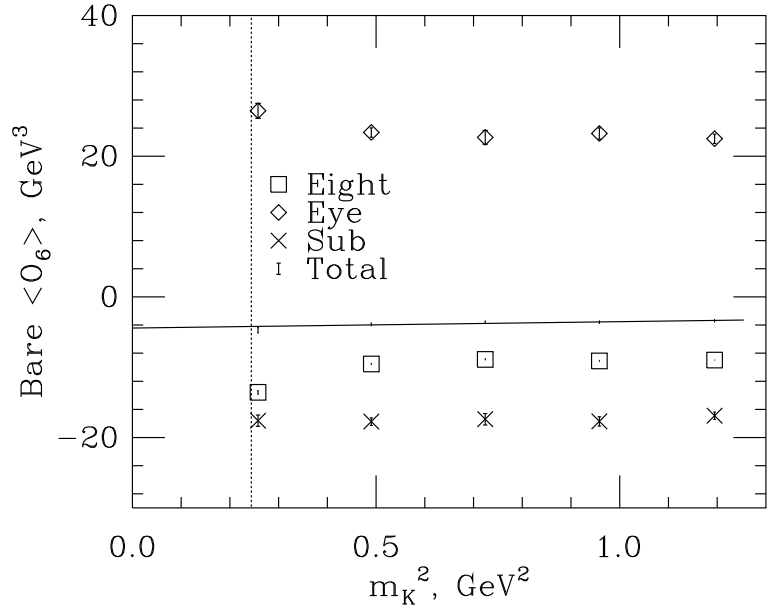}}
\end{center}
\caption{Three contributions to $\langle O_6\rangle$ matrix element:
``eight'' (boxes), ``eye'' (diamonds) and ``subtraction'' (crosses).
These data represent bare operators for the dynamical ensemble.
The fit is done for the sum total of all contributions. All errors
were combined by jackknife.}
\label{fig:O6}
\end{figure} 

\subsection{Problems with perturbative matching}

The value of $\varepsilon '/\varepsilon$ depends on a number of 
subtle cancellations
between matrix elements. In particular, $O_6$ and $O_8$
have been so far considered the most important operators
whose contributions have opposite signs and almost cancel. Furthermore,
matrix element of individual operators contain three main components 
(``eights'', ``eyes'',
and ``subtractions''), which again conspire to almost cancel each other
(see Fig.~\ref{fig:O6}). 
Thus $\varepsilon '/\varepsilon$ is extremely sensitive to each of these 
components, and in particular to their matching. 

\begin{table}[tbh]
\caption{$\langle O_6\rangle$ in arbitrary units with one-loop perturbative
matching using two values of $q^*$ for $Q_1$ ensemble. For comparison, 
the results with no matching (``bare'') are given.}
\label{tab:O6pert}
\begin{tabular}{l|ccccc}
\tableline\tableline
Quark mass & 0.01 & 0.02 & 0.03 & 0.04 & 0.05 \\
\tableline
$q^* =1/a$ & 
$0.1 \pm 1.2 $ &
$-0.9 \pm 0.4$ &
$-1.2 \pm 0.2$ &
$-1.6 \pm 0.3$ &
$-1.1 \pm 0.2$    \\
$q^*=\pi /a$ &
$-13.1 \pm 1.8$ &
$ -9.0 \pm 0.5$ &
$ -7.1 \pm 0.3$ &
$ -6.3 \pm 0.5$ &
$ -4.6 \pm 0.5$ \\
Bare &
$-55.6 \pm 5.0$ &
$-35.4 \pm 1.5$ &
$-27.0 \pm 0.9$ &
$-22.3 \pm 1.4$ &
$-16.4 \pm 1.5$ \\
\tableline\tableline
\end{tabular}
\end{table}

Consider fermion contractions with operators such as\footnote{
We apologize for slightly confusing notation:
we use the same symbols here for
operators as in the Appendix for types of contractions.}
\begin{eqnarray}
\label{eq:O6ops1}
(PP)_{EU} & = & (\overline{s} \gamma_5 \otimes \xi_5 u) 
(\overline{u} \gamma_5 \otimes \xi_5 d) \\
(SS)_{IU} & = & (\overline{s} \openone \otimes \openone d) 
(\overline{d} \openone \otimes \openone d) \\
\label{eq:O6ops3}
(PS)_{A2U} & = & (\overline{s} \gamma_5 \otimes \xi_5 d) 
(\overline{d} \openone \otimes \openone d) ,
\end{eqnarray}
which are main parts of, correspondingly, ``eight'', ``eye'' and 
``subtraction''
components of $O_6$ and $O_8$ (see the Appendix). The 
finite renormalization coefficients for these operators
have been  computed in Ref.~\cite{PatelSharpe}. 
The diagonal coefficients are very large, so the
corresponding one-loop corrections are in the 
neighborhood of $-100\%$. In addition, they strongly depend
on which $q^*$ is used (refer to Table~\ref{tab:O6pert}).
Thus perturbation theory fails in reliably 
matching the operators in Eqs.~(\ref{eq:O6ops1}--\ref{eq:O6ops3}). 

The finite coefficients for other (subdominant)
operators, for example
$(PP)_{EF}$, $(SS)_{EU}$ and $(SS)_{EF}$,
are not known for formulation with gauge-invariant 
operators\footnote{Patel and Sharpe~\cite{PatelSharpe}
have computed corrections for gauge-noninvariant operators.
Operators in Eqs.~(\ref{eq:O6ops1})--(\ref{eq:O6ops3})
have zero distances, so the corrections are the same for
gauge invariant and non-invariant operators.
Renormalization of other operators
(those having non-zero distances) differs from the 
gauge-noninvariant operators.}.
For illustration purposes,
in Table~\ref{tab:O6pert} we have used coefficients for gauge
non-invariant operators computed in Ref.~\cite{PatelSharpe}, but 
strictly speaking this is not justified. 

To summarize, perturbative matching does not work and
some of the coefficients are even poorly known. A solution
would be to use a non-perturbative matching procedure, such as
described by Donini {\it et al.}~\cite{non-pert}. We have not completed
this procedure. Nevertheless, can we say anything
about $\varepsilon '/\varepsilon$ at this moment?

\subsection{Partially nonperturbative matching}

\label{sec:ansatz}

As a temporary solution, we have adopted a partially
nonperturbative operator matching procedure, which makes use of
bilinear renormalization coefficients $Z_P$ and $Z_S$.
We compute the latter~\cite{PKbil}
following the non-perturbative method suggested by Martinelli 
{\it et al.}~\cite{martinelli}. Namely we study the inverse
of the ensemble-averaged quark propagator at large off-shell momenta
in a fixed (Landau) gauge. 
An estimate of the
renormalization of four-fermion operators can be obtained as follows. 

\begin{figure}[htb]
\begin{center}
\leavevmode
\centerline{\epsfysize=8cm \epsfbox{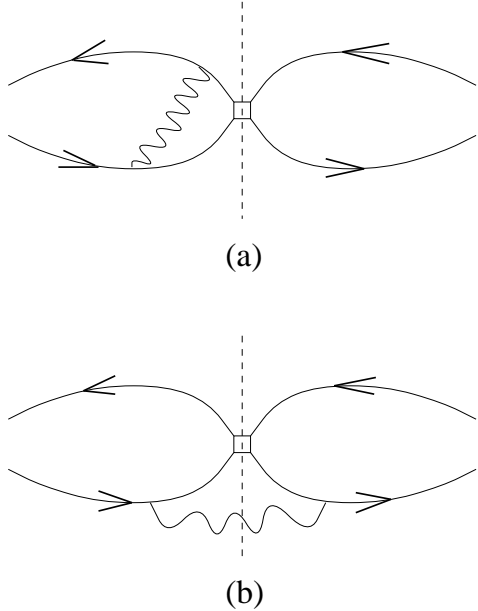}}
\end{center}
\caption{Example of one loop diagrams arising in 
renormalization of four-fermion operators: in type (a) no propagator
crosses the axis, and type (b) includes the rest of the diagrams.}
\label{Opp}
\end{figure} 

Consider
renormalization of the pseudoscalar--pseudoscalar operator in
Eq.~(\ref{eq:O6ops1}). 
At one-loop level, the diagonal renormalization coefficient 
$C_{PP}$ (involving diagrams shown in Fig.~\ref{Opp}) 
is almost equal to twice the pseudoscalar bilinear correction $C_P$. 
This suggests that, at least at one-loop level,
the renormalization of $(PP)_{EU}$ comes mostly from diagrams
in which no gluon propagator crosses the vertical axis of the diagram
(for example, diagram $(a)$ in Fig.~\ref{Opp}), and very little
from the rest of the diagrams
(such as diagram $(b)$ in Fig.~\ref{Opp}). In other words, the
renormalization of $(PP)_{EU}$ would be identical to 
the renormalization of product of two pseudoscalar bilinears,
were it not for the diagrams of type $(b)$, which give a subdominant
contribution. Mathematically, 
$$
(PP)_{EU}^{\mathrm{cont}} = (PP)_{EU}^{\mathrm{latt}}\; Z_{PP} + ... \, ,
$$
with
\begin{equation}
Z_{PP} = Z_P^2 (1 + \frac{g^2}{16\pi^2} \widetilde{C_{PP}} + O(g^4))\, ,
\label{eq:Zpp}
\end{equation}
\begin{equation}
Z_P = 1 + \frac{g^2}{16\pi^2} C_P + O(g^4)\, ,
\label{eq:Zp}
\end{equation}
and dots indicate mixing with other operators (non-diagonal part).
The factor $\widetilde{C_{PP}} \equiv C_{PP} - 2 C_P$ contains
diagrams of type $(b)$ in Fig.~\ref{Opp} and is quite small.

\begin{figure}[htbp]
\begin{center}
\leavevmode
\centerline{\epsfysize=14cm \epsfbox{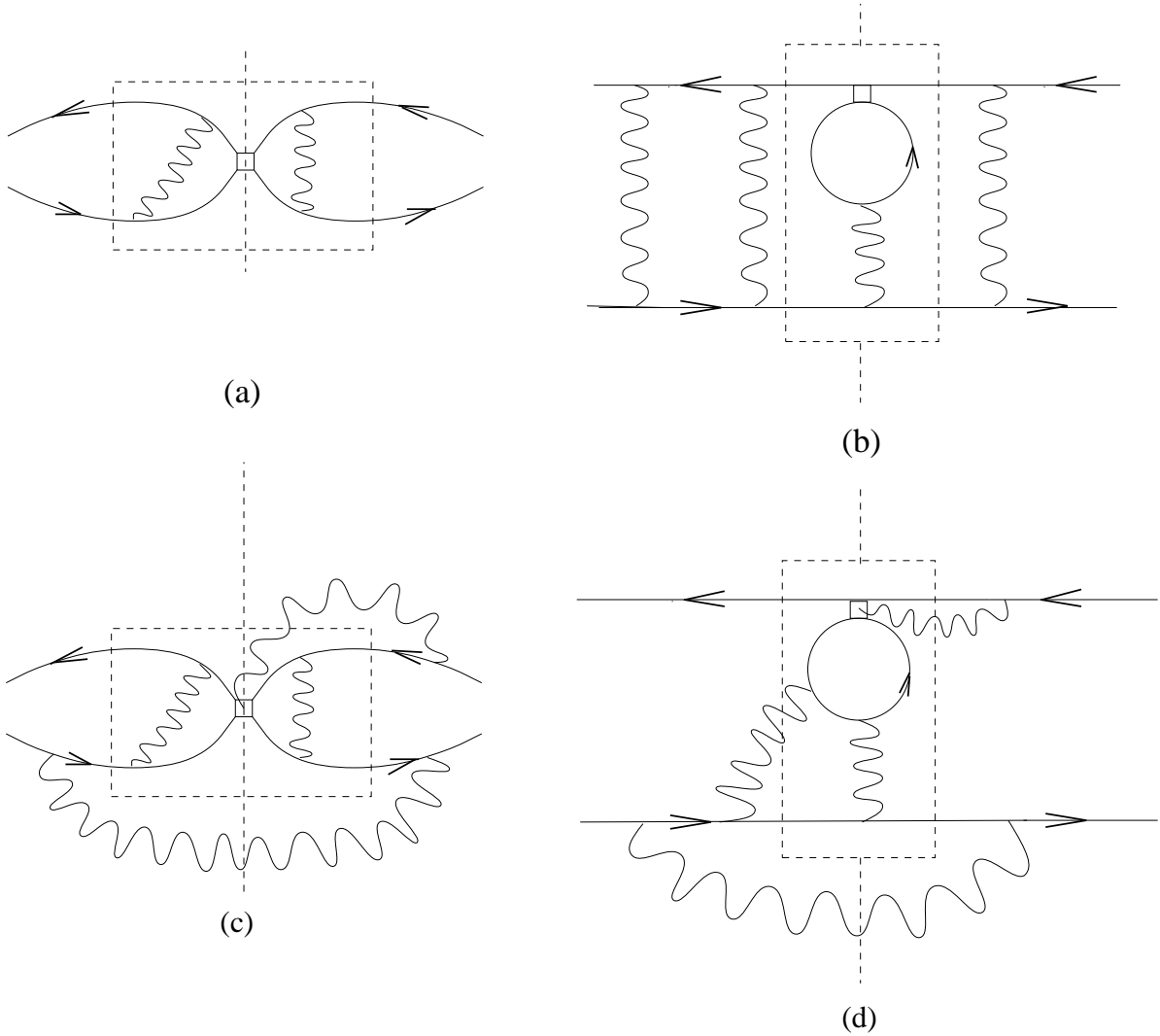}}
\end{center}
\caption{Example of four kinds of diagrams with arbitrary number of loops
arising in renormalization 
of four-fermion operators: in (a) and (b) no propagator
crosses the box or the axis; (c) and (d) exemplify the
rest of the diagrams. The rectangular drawn in dotted line in (b)
corresponds to operator structure $PP_{EU}$.}
\label{higher-order}
\end{figure}

In order to proceed, it may be reasonable to {\bf assume} that the same 
holds at all orders in perturbation
theory, namely the diagrams of type $(c)$ in Fig.~\ref{higher-order} give
subdominant contribution compared to type $(a)$ of the same
Figure. This assumption should be verified
separately by performing non-perturbative renormalization procedure
for four-fermion operators. If this ansatz is true, we can substitute
the non-perturbative value of $Z_P$ into Eq.~(\ref{eq:Zpp}) instead
of using the perturbative expression from Eq.~(\ref{eq:Zp}).
Thus a partially nonperturbative estimate of $(PP)_U^{\mathrm cont}$
is obtained. This procedure is quite similar to the tadpole
improvement idea: the bulk of diagonal renormalization is
calculated non-perturbatively, while the rest is reliably computed
in perturbation theory.  
Analogously we obtain diagonal renormalization
of operators $(SS)_{IU}$ and $(PS)_{A2U}$ by using
$Z_{SS} = Z_S^2(1+\frac{g^2}{16\pi^2} \widetilde{C_{SS}} + O(g^4))$ and 
$Z_{PS} = Z_S Z_P(1+\frac{g^2}{16\pi^2} \widetilde{C_{PS}} + O(g^4))$.
We note that $Z_P \neq Z_S$, even though they are equal in perturbation
theory. We match operators at the scale $q^*=1/a$ and use the
continuum two-loop anomalous dimension to evolve to $\mu =2$ GeV.

Unfortunately, the above procedure does not solve completely the problem 
of operator renormalization, since it deals only with diagonal 
renormalization of the zero-distance operators in
Eqs.~(\ref{eq:O6ops1}---\ref{eq:O6ops3}). Even though these operators
are dominant in contributing to $\varepsilon '/\varepsilon$, other
operators (such as $(SS)_{EU}$ and $(PP)_{EF}$)
can be important due to mixing with the dominant ones.
The mixing coefficients for these operators are not known 
even in perturbation theory. For a reasonable estimate we use
the coefficients
obtained for gauge non-invariant operator mixing~\cite{PatelSharpe}.

Secondly, since renormalization of operators $(PP)_{EU}$, $(SS)_{IU}$ 
and $(PS)_{A2U}$ is dramatic\footnote{For example, at $m_q=0.01$ and 
$\mu =2$ GeV for $Q_1$ ensemble we obtain $Z_{PP} = 0.055 \pm 0.007$, 
$Z_{PS} = 0.088 \pm 0.007$ and $Z_{SS} = 0.142 \pm 0.010$.}, their 
influence on other operators 
through non-diagonal mixing is ambiguous at one-loop order, 
even if the mixing coefficients are known.
The ambiguity is due to higher
order diagrams (for example, those shown in Fig.~\ref{higher-order}). 
In order to partially resum them
we use operators $(PP)_{EU}$, $(SS)_{IU}$ and $(PS)_{A2U}$ 
multiplied by factors $Z_P^2$, $Z_S^2$ and $Z_P Z_S$, correspondingly,
whenever they appear in non-diagonal mixing with other operators
\footnote{
A completely analogous scheme was used for mixing of $O_6$ with $O_2$ 
through penguins when evaluating $\Re A_0$.}. 
This is equivalent to evaluating the diagrams of type ($a$) and ($b$)
in Fig.~\ref{higher-order} at all orders, but ignoring the rest
of the diagrams (such as diagrams ($c$) and ($d$) in Fig~\ref{higher-order})
at all orders higher than first.
To estimate a possible error in this procedure
we compare with a simpler one, whereby bare operators
are used in non-diagonal corrections (i.e. we apply strictly one-loop 
renormalization).
The difference in $\varepsilon '/\varepsilon$ between these two approaches
is of the same order or even less than the error due to uncertainties in 
determination
of $Z_P$ and $Z_S$ (see Tables~\ref{tab:epsp1} and~\ref{tab:epsp2}). 

\begin{figure}[htbp]
\begin{center}
\leavevmode
\centerline{\epsfysize=8cm \epsfbox{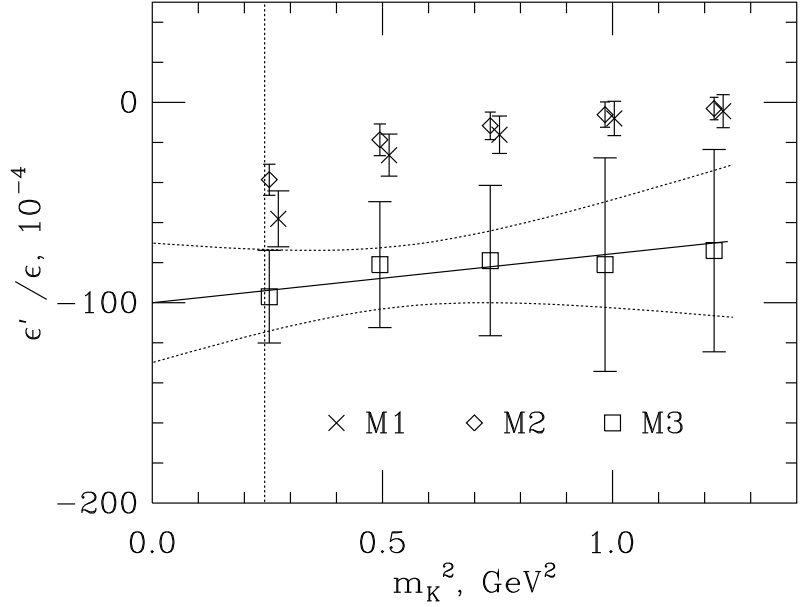}}
\end{center}
\caption{A rough estimate of $\varepsilon '/\varepsilon$ 
for $Q_1$ ensemble using the partially-nonperturbative procedure
described in text. Three sets of points correspond to 
using experimental $\Re A_0$ and $\Re A_2$ in 
Eq.~(\ref{eq:epsp}) (crosses),
using our $\Re A_0$ but experimental $\omega$ (diamonds), or using 
$\Re A_0$ and $\Re A_2$ obtained from our calculations (squares).
All other details are the same as in Table~\ref{tab:epsp1}.
The shown error is a combination of the statistical error, 
a matching error coming from uncertainties in the determination of $Z_P$ 
and $Z_S$, and an uncertainty in non-diagonal mixing of
subdominant operators.}
\label{epsp}
\end{figure} 

\clearpage

\section{Estimates of $\varepsilon '/\varepsilon$}

\label{sec:epsp_res}

Within the procedure outlined in the previous section we have found that 
$\langle O_6\rangle$ has a different sign from the expected one.
This translates into a negative or very slightly positive value of 
$\varepsilon '/\varepsilon$ (Tables~\ref{tab:epsp1} and~\ref{tab:epsp2}
and Fig.~\ref{epsp}). 

If the assumptions about the subdominant diagrams made in the previous 
section are valid, our results would contradict the present
experimental results, which favor a positive value of 
$\varepsilon '/\varepsilon$. They would also change the existing
theoretical picture~\cite{buras} due to the change of sign of 
$\langle O_6\rangle$.

Finite volume and quenching effects were found small
compared to noise. 
The main uncertainty in $\varepsilon '/\varepsilon$ 
comes from operator matching, diagonal and non-diagonal. 
For diagonal matching the uncertainty comes from (1) errors in the
determination of $Z_P$ and $Z_S$ non-perturbatively and from
(2) unknown degree of validity of our ansatz in Sec.~\ref{sec:ansatz}. 
For non-diagonal matching, the error is due to (3) unknown
non-diagonal coefficients in mixing matrix and (4) ambiguity 
of accounting higher-order corrections. 
The error (1), as well as the statistical error, is quoted in 
Tables~\ref{tab:epsp1} and~\ref{tab:epsp2}. The size
of the error (4) can be judged by the spread
in $\varepsilon '/\varepsilon$ between two different
approaches to higher-order corrections (strictly one-loop and partial
resummation), also presented in
Tables~\ref{tab:epsp1} and~\ref{tab:epsp2}. The error (3) is likely
to be of the same order as the error (4).
The error (2) is uncontrolled at this point, since it 
is difficult to rigorously check our assumption made 
in Sec.~\ref{sec:ansatz}.  In Fig.~\ref{epsp} we combine
the statistical error with errors (1) and (4) in quadrature.

The uncertainty due to operator matching is common to any method
to compute the relevant matrix elements on the lattice (at least, with 
staggered fermions). In addition, our method has an inherent
uncertainty due to dropping the higher order chiral terms.
Lattice spacing dependence of $\varepsilon '/\varepsilon$ is
unclear at this point, but it may be significant.

We note that there are several ways to compute $\varepsilon '/\varepsilon$.
One can use the experimental values of $\Re A_0$ and $\Re A_2$ in 
Eq.~(\ref{eq:epsp}), or one can use the values obtained on the lattice.
One can also adopt an intermediate strategy of using the experimental
amplitude ratio $\omega$ and computed $\Re A_0$. When the higher-order
chiral corrections are taken into account and the continuum limit is taken
(so that $\omega = 22$),
these three methods should converge. At this point any of them
can be used, and we compare them in Tables~\ref{tab:epsp1} 
and~\ref{tab:epsp2}.

In view of the issues raised above, $\varepsilon '/\varepsilon$ is an 
extremely fragile quantity. The rough estimates in Tables~\ref{tab:epsp1} 
and~\ref{tab:epsp2} and Fig.~\ref{epsp} should be used with extreme 
caution. 

\section{Conclusion}

\label{sec:conclusion}

We have presented in detail the setup of our calculation of
hadronic matrix elements of all operator in the basis
defined in Eqs.~(\ref{eq:ops1}---\ref{eq:ops10}).
We have obtained statistically significant data for all operators.
Based on these data we make theoretical estimates of $\Re A_0$ and
$\Re A_2$ amplitudes as well as $\varepsilon '/\varepsilon$.

Simulation results show that the enhancement of the $\Delta I=1/2$ 
transition is roughly consistent with the experimental findings. 
However, the uncertainty due to higher order
chiral terms is very large. If these terms are calculated
in the future, a more definite prediction for physical amplitudes
can be obtained using our present data for matrix elements.
Simulations should be repeated at a few more values of $\beta$ 
in the future in order to take the continuum limit. 

Calculation of $\varepsilon '/\varepsilon$ is further complicated 
by the failure
of perturbation theory in operator matching. We give our
crude estimates, but in order to achieve real progress 
the full nonperturbative matching procedure should be performed.

We appreciate L. Venkataraman's help in developing 
CRAY-T3E software. 
We thank the Ohio Supercomputing Center and National Energy Research 
Scientific Computing Center (NERSC) for the CRAY-T3E time. We thank 
Columbia University group for  access to their dynamical 
configurations.  

\appendix

\section*{Explicit expressions for fermion contractions}

\subsection{Quark operators}

We work in the two flavor traces formalism when calculating
contractions with four-fermion operators: for each 
contraction separately the operators are rendered in the form 
(if necessary, by Fierz transformation) of two bilinears with the 
flavor flow in the form of a product of two flavor traces. To be more 
precise, for ``eight'' contractions the operators  are rendered
in the form $(\overline{s}\Gamma u)(\overline{u}\Gamma d)$,
while for the ``eye'' and ``annihilation'' contractions
the appropriate form is $(\overline{s}\Gamma d)
(\overline{q}\Gamma q)$. This is done in the continuum, before
assigning the staggered fermion flavour.

The operator transcription in flavor space for staggered fermions is now 
standard~\cite{WeakME}, and we give it here for completeness. 
The Goldstone bosons have spin-flavor
structure $\gamma_5 \otimes \xi_5$. The flavor structure of the operators
is defined by requiring non-vanishing of the flavor traces,
and so it depends on the contraction type: the flavor structure is $\xi_5$
in ``eights'' and two-point functions, $\openone$ in ``eyes''
and ``subtractions''. In ``annihilation'' contractions
the flavour structure is $\openone$ for the bilinear in the quark loop trace 
and $\xi_5$ for the one involved in the external trace.

Either one or two color traces may be appropriate for a particular
contraction with a given operator (see the next Appendix section
for details). In one trace contractions (type ``F'' for ``fierzed'') the
color flow is exchanged between the bilinears, while in 
two trace contractions (type ``U'' for ``unfierzed'') the color
flow is contained within each bilinear so that the contraction
is the product of two color traces. In either contraction type,
when the distance between staggered fermion fields 
being color-connected is non-zero, a gauge connector is inserted
in the gauge-invariant fashion. The connector is computed as the average 
of products of gauge links along all shortest paths connecting the 
two sites. We also implement tadpole improvement
by dividing each link in every gauge connector by $u_0 = (1/3 \mathrm{Tr}
(U_P))^{1/4}$, where $U_P$ is the average plaquette value. 

\subsection{Sources and contractions}

\label{sources}

We use local $U(1)$ pseudofermion wall sources. Explicitly,
we set up a field of $U(1)$ phases $\xi_\alpha({\mathbf x};t_0)$
($|\xi_\alpha | = 1$) for each color and each site at a given timeslice 
$t_0$, which are chosen at random and satisfy
\begin{equation}
\label{eq:noise}
\langle\xi_\alpha^*({\mathbf x};t_0)\; \xi_\beta ({\mathbf y};t_0)\rangle = 
\delta_{\alpha ,\beta }\; \delta_{{\mathbf x},{\mathbf y}}. 
\end{equation}
(Boldface characters designate spatial parts of the 4-vector
with the same name.) 
We proceed to explain how this setup works in the case of the
two-point function calculation, with trivial generalization to ``eight''
and ``annihilation'' contractions.

Consider the propagator from a wall at $t_0=0$ in a given background
gauge configuration, computed by
inverting the equation
\begin{equation}
\label{eq:prop}
(\rlap{\,/}D +m)^{\alpha\beta}_{x\;y}\; \chi_{\beta}(y) = 
\xi_\alpha ({\mathbf x};0)\delta_{x_4,0}.
\end{equation}
This is equivalent to computing
\begin{equation}
\chi_\beta (y) = \sum_{\mathbf x} \, \xi_\alpha ({\mathbf x};0)
G_{\beta\alpha}(y;{\mathbf x},0)\, ,
\end{equation}
where $G(y;x)$ is the propagator from 4-point x to 4-point y. 
For staggered fermions description we label
the fields by hypercube 
index $h$ and the hypercube corner indices $A_\mu \in \{0,1\}^4$ instead
of $y$. The two-point function is constructed as follows:
\begin{equation}
\label{eq:2p-function}
{\mathrm TP} \propto \sum_{h,A} \chi^*_{\alpha} (h,A) 
U_{\alpha\beta}(h,A,A+\Delta ) \chi_{\beta}(h,A+\Delta ) \;
\phi (A)\;(-1)^A ,
\end{equation}
where $\phi (A)$ and $\Delta_\mu$ are phases and distances
appropriate for a given staggered fermion operator
\footnote{For a given bilinear with spin-flavor structure 
  $\Gamma_S\otimes\Gamma_F$, these are determined as follows: 
\mbox{$\Delta_\mu = |S_\mu-F_\mu|^2$} and $\phi (A) = \frac{1}{4} {\mathrm Tr} 
(\Gamma_A^\dag\,\Gamma_S\,\Gamma_{A+\Delta}\,\Gamma_F^\dag )$, 
where $S_\mu$ and $F_\mu$ are spin and flavor vectors
such that $\Gamma_S = \gamma_1^{S_1}\gamma_2^{S_2}\gamma_3^{S_3}\gamma_4^{S_4}$
and $\Gamma_F = \gamma_1^{F_1}\gamma_2^{F_2}\gamma_3^{F_3}\gamma_4^{F_4}$,
and $\Gamma_A = \gamma_1^{A_1}\gamma_2^{A_2}\gamma_3^{A_3}\gamma_4^{A_4}$.}
, $U(h,A,A+\Delta )$ is
the appropriate gauge connector (see below), modulo 2 summation is implied
for hypercube indices $A$, and $h$ runs over all hypercubes
in a given timeslice $t$ where the operator is inserted. The factor $(-1)^A$ 
takes into account that for staggered fermions 
$G(x;y) = G^\dag (y;x) (-1)^x(-1)^y$.
Equation~(\ref{eq:2p-function}) corresponds to 
\begin{equation}
{\mathrm TP} \propto \sum_{\mathbf x,y,z} G_{\alpha\beta}(z,y) \;\Gamma \;G_{\beta\gamma}(y,x)\; (-1)^z 
\;\xi^*_\alpha (z)\xi_\gamma(x) \, ,
\end{equation}
where $\Gamma$ is used for simplicity to show the appropriate operator
structure. The summation over ${\mathbf x}$ and ${\mathbf z}$ 
over the entire spatial volume averages over the noise, so the last equation
is equivalent to
\begin{equation}
\label{eq:123}
{\mathrm TP} \propto \sum_{\mathbf x,y}{\mathrm tr}\; G(x,y) \,\Gamma\, G(y,x)\; (-1)^x .
\end{equation}
Therefore, using the pseudofermion wall source is equivalent to 
summation of contractions obtained with independent local delta-function 
sources.
Note that the factor $(-1)^x$ and zero distance in the staggered fermions 
language are equivalent to spin-flavor structure $\gamma_5 \otimes \xi_5$. 
This means the source creates pseudoscalar mesons at rest, which 
includes Goldstone bosons.
Strictly speaking, this source also creates mesons with spin-flavor
structure $\gamma_5\gamma_4 \otimes \xi_5\xi_4$, since it is defined 
only on one timeslice. However, as explained in the first footnote
in Section~2.3 of Ref.~\cite{WeakME}, these states do not 
contribute. 

We have used one copy of pseudofermion sources per configuration.

Analogously, we construct the pion sink at time $T$ by using 
another set of $U(1)$ random noise 
($\langle\xi_\alpha^*({\mathbf x};T)\; \xi_\beta ({\mathbf y};T)
\rangle = \delta_{\alpha ,\beta }\; \delta_{{\mathbf x},{\mathbf y}}$, 
$|\xi |=1$). The propagator $\Phi$ is computed as follows:
\begin{equation}
(\rlap{\,/}D +m)^{\alpha\beta}_{x\;y}\; \Phi_{\beta}(y) = 
\xi_\alpha ({\mathbf x};T)\delta_{x_4,T}. 
\end{equation}

Suppose $\Delta_1$, $\Delta_2 \in \{0,1\}^4$ and $\phi_1 (A)$, $\phi_2 (A)$
are distances and phases of the two staggered fermion bilinears making
up a given four-fermion operator. 
The expression for the ``eight'' contraction (Fig.~\ref{diagrams}a) with two
color traces (``U'' type) is given by
\begin{eqnarray}
\label{eq:eight}
{\mathrm E_U} \propto \sum_{h,A,B} & 
\chi^*_{\alpha} (h,A) U_{\alpha\beta}(h,A,A+\Delta_1 ) 
\chi_{\beta }(h,A+\Delta_1 ) \;\phi_1 (A)\;(-1)^A \nonumber \\
& \times \Phi^*_{\rho} (h,B) U_{\rho\sigma}(h,B,B+\Delta_2 ) 
\Phi_{\sigma}(h,B+\Delta_2 ) \;\phi_2 (B)\;(-1)^B ,
\end{eqnarray}
up to various normalization factors which cancel in the $B$ ratio. 
In this expression $A$, $B \in \{0,1\}^4$ run over 16 hypercube corners 
(modulo 2 summation is implied for these indices). The hypercube
index $h$, as before, runs over the entire spatial volume of the
timeslice $t$ of the operator insertion. The gauge connector
$U (h,A,B )$ is the identity matrix when $A = B$, otherwise it is the 
average of products of gauge links in the given configuration along all 
shortest paths from $A$ to $B$ in a given hypercube $h$. The 
expression~(\ref{eq:eight}), as well as all other contractions,
is computed for each background gauge configuration and
is subject to averaging over the configurations. 
(Whenever several contractions are combined in a single
quantity, such as a $B$ ratio, we use jackknife to estimate
the statistical error).

The expression for one color trace (``F'' type) contraction is similar:
\begin{eqnarray}
{\mathrm E_F} \propto \sum_{h,A,B}  &
\chi^*_{\alpha} (h,A) U_{\alpha\beta}(h,A,B+\Delta_2 ) 
\chi_{\sigma}(h,A+\Delta_1 ) \;\phi_1 (A)\;(-1)^A \nonumber \\
& \times \Phi^*_{\rho} (h,B) U_{\rho\sigma}(h,B,A+\Delta_1 ) 
\Phi_{\beta }(h,B+\Delta_2 ) \;\phi_2 (B)\;(-1)^B ,
\end{eqnarray}

For ``eye'' and ``subtraction'' diagrams (Fig.~\ref{diagrams}b
and~\ref{diagrams}d) the source setup is a little more
involved, since the kaon and pion are directly connected
by a propagator. In order to construct a wall source we need to compute 
the product
$$
\psi (y) = \sum_{\mathbf{x}} G({\mathbf y},t; {\mathbf x},T) \cdot 
G({\mathbf x},T;{\mathbf 0},0) (-1)^x.
$$
In order to avoid computing propagators from every point $\mathbf{x}$ 
at the timeslice $T$, we first compute propagator
$G({\mathbf x},T;{\mathbf 0},0)$, cut out the timeslice $T$ and use it as the 
source for calculating the propagator to $({\mathbf y},t)$. This amounts
to inverting equation 
\begin{equation}
(\rlap{\,/}D +m)_{x\;y}^{\alpha\beta}\; \psi_{\beta}(y) = 
\chi_{\alpha} (x) \; \delta_{(x_4,T)} (-1)^x\, ,
\end{equation}
where $\chi_\alpha (x)$ is the propagator from the wall source 
at $t_0=0$ defined in 
Eq.~(\ref{eq:prop}). We use the following expression for evaluating the
``subtraction'' diagram:
\begin{equation}
\label{eq:sub_diag}
{\mathrm S} \propto \sum_{h,A} \chi^*_{\alpha} (h,A) 
U_{\alpha\beta}(h,A,A+\Delta )\psi_{\beta} (h,A+\Delta ) \;
\phi (A)\;(-1)^A ,
\end{equation}
Again, averaging over the noise leaves only local connections in both
sources, so in the continuum language we get:
\begin{equation}
{\mathrm S} \propto \sum_{\mathbf x,y,z} {\mathrm tr} \;
G({\mathbf x},0;{\mathbf z},t) \;
\Gamma \; G({\mathbf z},t;{\mathbf y},T) \;
G({\mathbf y},T;{\mathbf x},0)\; (-1)^x (-1)^y\, .
\end{equation}
(In fact, we are mostly interested in subtracting the operator
$\overline{s} \openone\otimes\openone d$,
so in Eq.~(\ref{eq:sub_diag}) $\Delta = \{0,0,0,0\}$ and $\phi (A)=1$.)

In order to efficiently compute fermion loops for ``eye'' and 
``annihilation'' diagrams (Fig.~\ref{diagrams}b and~\ref{diagrams}c), 
we use $U(1)$ noise copies $\zeta^{(i)}$, $i=1,\dots ,N$, 
at every point in space-time. We compute $\eta^{(i)}$ by inverting
$(\rlap{\,/}D +m)\eta^{(i)} = \zeta^{(i)}$. 
It is easy to convince oneself that the propagator from $y$ to $x$ equals
\begin{equation}
G(x;y) = \langle \eta_x \zeta_y^*\rangle.
\end{equation}
In practice we average over $N=10$ noise copies.
This includes 2 or 4 copies of the lattice in time extension, 
so the real number of noise copies is 20 or 40, with another 
factor of 3 for color. The efficiency of this method is crucial for
obtaining good statistical precision.

The expression for 
``U'' and ``F'' type ``eye'' diagrams are as follows:
\begin{eqnarray}
{\mathrm I_U} & \propto  \displaystyle\sum_{h,A,B} &
\chi^*_{\alpha} (h,A) U_{\alpha\beta}(h,A,A+\Delta_1 ) 
\psi_{\beta}(h,A+\Delta_1 ) \;\phi_1 (A)\;(-1)^A \nonumber \\
\times & \displaystyle\frac{1}{N}\sum_{i=1}^N & 
\zeta^{(i)*}_\rho (h,B)U_{\rho\sigma}(h,B,B+\Delta_2) 
\eta^{(i)}_\sigma (h,B+\Delta_2 ) \;\phi_2 (B)\;(-1)^B , \\
{\mathrm I_F} & \propto  \displaystyle\sum_{h,A,B} &
\chi^*_{\alpha} (h,A) U_{\alpha\sigma}(h,A,B+\Delta_2 ) 
\psi_{\beta}(h,A+\Delta_1 ) \;\phi_1 (A)\;(-1)^A \nonumber \\
\times & \displaystyle\frac{1}{N}\sum_{i=1}^N & 
\zeta^{(i)*}_\rho (h,B)U_{\rho\beta}(h,B,A+\Delta_1) 
\eta^{(i)}_\sigma (h,B+\Delta_2 ) \;\phi_2 (B)\;(-1)^B .
\end{eqnarray}

The computation of ``annihilation'' diagrams (Fig.~\ref{diagrams}c)
is similar to the two-point function, except the fermion loop is 
added and the derivative with respect to the quark mass 
difference $m_d-m_s$ 
is inserted in turn in every strange quark propagator. 
Derivatives of the propagators are given by inverting equations
\begin{eqnarray}
(\rlap{\,/}D +m)\chi ' & = & \chi , \\
(\rlap{\,/}D +m)\eta^{'(i)} & = & \eta^{(i)} .
\end{eqnarray}
We have, therefore, four kinds of ``annihilation'' contractions,
which should be combined in an appropriate way for each operator
depending on the quark flavor structure (this is spelled out in the next
Appendix section):
\begin{eqnarray}
  {\mathrm A_{1U}} & \propto \displaystyle\sum_{h,A,B} &
\chi^{'*}_{\alpha} (h,A) U_{\alpha\beta}(h,A,A+\Delta_1 ) 
\chi_{\beta }(h,A+\Delta_1 ) \;\phi_1 (A)\;(-1)^A \nonumber \\
\times & \displaystyle\frac{1}{N}\sum_{i=1}^N &
\zeta^{(i)*}_\rho (h,B)U_{\rho\sigma}(h,B,B+\Delta_2) 
\eta^{(i)}_\sigma (h,B+\Delta_2 ) \;\phi_2 (B)\;(-1)^B , \\
  {\mathrm A_{1F}} & \propto \displaystyle\sum_{h,A,B} &
\chi^{'*}_{\alpha} (h,A) U_{\alpha\sigma}(h,A,B+\Delta_2 ) 
\chi_{\beta}(h,A+\Delta_1 ) \;\phi_1 (A)\;(-1)^A 
\nonumber \nolinebreak\\
\times & \displaystyle\frac{1}{N}\sum_{i=1}^N &
\zeta^{(i)*}_\rho (h,B)U_{\rho\beta}(h,B,A+\Delta_1) 
\eta^{(i)}_\sigma (h,B+\Delta_2 ) \;\phi_2 (B)\;(-1)^B , \\
  {\mathrm A_{2U}} & \propto \displaystyle\sum_{h,A,B} &
\chi^{*}_{\alpha} (h,A) U_{\alpha\beta}(h,A,A+\Delta_1 ) 
\chi_{\beta }(h,A+\Delta_1 ) \;\phi_1 (A)\;(-1)^A \nonumber \\
\times & \displaystyle\frac{1}{N}\sum_{i=1}^N &
\zeta^{(i)*}_\rho (h,B)U_{\rho\sigma}(h,B,B+\Delta_2) 
\eta^{'(i)}_\sigma (h,B+\Delta_2 ) \;\phi_2 (B)\;(-1)^B , \\
  {\mathrm A_{2F}} & \propto \displaystyle\sum_{h,A,B} &
\chi^{*}_{\alpha} (h,A) U_{\alpha\sigma}(h,A,B+\Delta_2 ) 
\chi_{\beta}(h,A+\Delta_1 ) \;\phi_1 (A)\;(-1)^A 
\nonumber \nolinebreak\\
\times & \displaystyle\frac{1}{N}\sum_{i=1}^N &
\zeta^{(i)*}_\rho (h,B)U_{\rho\beta}(h,B,A+\Delta_1) 
\eta^{'(i)}_\sigma (h,B+\Delta_2 ) \;\phi_2 (B)\;(-1)^B .
\end{eqnarray}

\section*{Explicit expressions for matrix elements in terms of
fermion contractions.}

Operators in Eqs.~(\ref{eq:ops1}---\ref{eq:ops10}) can be decomposed
into $I=0$ and $I=2$ parts, which contribute, correspondingly, 
to $\Delta I=1/2$ and $\Delta I=3/2$ transitions. Here we give the 
expressions for these parts for completeness, since $\Re A_0$, 
$\Re A_2$ and $\varepsilon '/\varepsilon$ are directly expressible
in terms of their matrix elements. The $I=0$ parts are given
as follows:
\begin{eqnarray}
O_1^{(0)} & = & \frac{2}{3} 
(\overline{s}\gamma_\mu (1-\gamma_5)d)(\overline{u}\gamma^\mu (1-\gamma_5)u)
-\frac{1}{3}(\overline{s}\gamma_\mu (1-\gamma_5)u)(\overline{u}\gamma^\mu 
(1-\gamma_5)d)  \nonumber \\
& + & \frac{1}{3}(\overline{s}\gamma_\mu (1-\gamma_5)d)
(\overline{d}\gamma^\mu (1-\gamma_5)d) \\
O_2^{(0)} & = & \frac{2}{3} 
(\overline{s}\gamma_\mu (1-\gamma_5)u)(\overline{u}\gamma^\mu (1-\gamma_5)d)
-\frac{1}{3}(\overline{s}\gamma_\mu (1-\gamma_5)d)(\overline{u}\gamma^\mu 
(1-\gamma_5)u) \nonumber \\
& + & \frac{1}{3}(\overline{s}\gamma_\mu (1-\gamma_5)d)
(\overline{d}\gamma^\mu (1-\gamma_5)d) \\
O_3^{(0)} & = & 
(\overline{s}\gamma_\mu (1-\gamma_5)d) \sum_{q=u,d,s}(\overline{q}\gamma^\mu (1-\gamma_5)q) \\
O_4^{(0)} & = & 
(\overline{s}_\alpha\gamma_\mu (1-\gamma_5)d_\beta ) \sum_{q=u,d,s}(\overline{q}_\beta
\gamma^\mu (1-\gamma_5)q_\alpha ) \\
O_5^{(0)} & = & 
(\overline{s}\gamma_\mu (1-\gamma_5)d) \sum_{q=u,d,s}
(\overline{q}\gamma^\mu (1+\gamma_5)q) \\
O_6^{(0)} & = & 
(\overline{s}_\alpha\gamma_\mu (1-\gamma_5)d_\beta ) \sum_{q=u,d,s}
(\overline{q}_\beta \gamma^\mu (1+\gamma_5)q_\alpha ) \\
O_7^{(0)} & = & \frac{1}{2} [
(\overline{s}\gamma_\mu (1-\gamma_5)d)(\overline{u}\gamma^\mu (1+\gamma_5)u)
-(\overline{s}\gamma_\mu (1-\gamma_5)u)(\overline{u}\gamma^\mu (1+\gamma_5)d)
\nonumber \\ & - & (\overline{s}\gamma_\mu (1-\gamma_5)d)
(\overline{s}\gamma^\mu (1+\gamma_5)s)
] \\
O_8^{(0)} & = & \frac{1}{2} [
(\overline{s}_\alpha\gamma_\mu (1-\gamma_5)d_\beta )
(\overline{u}_\beta\gamma^\mu (1+\gamma_5)u_\alpha )
-(\overline{s}_\alpha\gamma_\mu (1-\gamma_5)u_\beta )
(\overline{u}_\beta\gamma^\mu (1+\gamma_5)d_\alpha ) \nonumber \\
& - & (\overline{s}_\alpha\gamma_\mu (1-\gamma_5)d_\beta )
(\overline{s}_\beta\gamma^\mu (1+\gamma_5)s_\alpha )
] \\
O_9^{(0)} & = & \frac{1}{2} [
(\overline{s}\gamma_\mu (1-\gamma_5)d)(\overline{u}\gamma^\mu (1-\gamma_5)u)
-(\overline{s}\gamma_\mu (1-\gamma_5)u)(\overline{u}\gamma^\mu (1-\gamma_5)d)
\nonumber \\ & - & (\overline{s}\gamma_\mu (1-\gamma_5)d)
(\overline{s}\gamma^\mu (1-\gamma_5)s)
] \\
O_{10}^{(0)} & = & \frac{1}{2} [
(\overline{s}\gamma_\mu (1-\gamma_5)u)(\overline{u}\gamma^\mu (1-\gamma_5)d)
-(\overline{s}\gamma_\mu (1-\gamma_5)d)(\overline{u}\gamma^\mu (1-\gamma_5)u)
\nonumber \\ & - & (\overline{s}\gamma_\mu (1-\gamma_5)d)
(\overline{s}\gamma^\mu (1-\gamma_5)s)
] 
\end{eqnarray}

Expressions for $I=2$ parts are as follows:

\begin{eqnarray}
O_1^{(2)} & = & O_2^{(2)} = \frac{2}{3}O_9^{(2)} = \frac{2}{3}O_{10}^{(2)}
= \frac{1}{3} [ 
(\overline{s}\gamma_\mu (1-\gamma_5)u)(\overline{u}\gamma^\mu (1-\gamma_5)d) 
\nonumber \\
& & +(\overline{s}\gamma_\mu (1-\gamma_5)d)(\overline{u}\gamma^\mu 
(1-\gamma_5)u) 
- (\overline{s}\gamma_\mu (1-\gamma_5)d)
(\overline{d}\gamma^\mu (1-\gamma_5)d)] \\
O_7^{(2)} & = & \frac{1}{2} [
(\overline{s}\gamma_\mu (1-\gamma_5)u)(\overline{u}\gamma^\mu (1+\gamma_5)d)
\nonumber \\
& & +(\overline{s}\gamma_\mu (1-\gamma_5)d)(\overline{u}\gamma^\mu 
(1+\gamma_5)u)
-  (\overline{s}\gamma_\mu (1-\gamma_5)d)
(\overline{d}\gamma^\mu (1+\gamma_5)d)] \\
O_8^{(2)} & = & \frac{1}{2} [
(\overline{s}_\alpha\gamma_\mu (1-\gamma_5)u_\beta )(\overline{u}_\beta
\gamma^\mu (1+\gamma_5)d_\alpha ) \nonumber \\
& &+(\overline{s}_\alpha\gamma_\mu (1-\gamma_5)d_\beta )(\overline{u}_\beta
\gamma^\mu (1+\gamma_5)u_\alpha ) 
-  (\overline{s}_\alpha\gamma_\mu (1-\gamma_5)d_\beta )
(\overline{d}_\beta\gamma^\mu (1+\gamma_5)d_\alpha )] \\
O_3^{(2)} & = & O_4^{(2)} = O_5^{(2)} = O_6^{(2)} = 0 
\end{eqnarray}
(Whenever the color indices are not shown, they are contracted
within each bilinear, i.e. there are two color traces.)

As mentioned in Sec.~\ref{sec:diag}, in order to compute
matrix elements of $I=0$ operators one needs to evaluate 
three types of diagrams: ``eight'' (Fig.~\ref{diagrams}a),
``eye'' (Fig.~\ref{diagrams}b) and ``annihilation''
(Fig.~\ref{diagrams}c). In the previous Appendix section we
have given detailed expressions for computation
of these contractions, given the spin-flavor structure.
Here we assign this structure to all contractions required for each
operator, i.e. we express each matrix element in terms of
contractions which were ``built'' in the previous section. 

Let us introduce some notation. Matrix element of the above
operators have three components:
\begin{equation}
\langle\pi^+\pi^-|O_i|K^0\rangle = (E_i + I_i - S\,(\,2\,m\alpha_i )\,  )\,
\frac{m_K^2-m_\pi^2}{(p_\pi \cdot p_K)f},
\end{equation}
where $m$ is the common quark mass for $s$, $d$ and $u$, and
\begin{equation}
\label{eq:alpha}
\alpha_i = \frac{A_i}{P}.
\end{equation}
Here $E_i$ and $I_i$ stand for ``eight and ``eye'' contractions
of the $\langle\pi^+|O_i|K^+\rangle$ matrix element, 
$A_i \sim \langle 0|O_i|K^0\rangle \,/\,(m_d-m_s)$ is the
``annihilation'' diagram, 
$S =\langle \pi^+|\overline{s}d|K^+\rangle$
is the ``subtraction'' diagram, and 
$P =\langle 0|\overline{s}\gamma_5 d|K^0\rangle$ is the two-point function.
We compute $\alpha_i$ by averaging the ratio in the right-hand side of
Eq.~(\ref{eq:alpha}) over a suitable time range. 

Detailed expressions for $E_i$, $I_i$ and $A_i$ 
are given below in terms of the basic contractions on the lattice.
We label basic contractions by two letters, each representing a bilinear.
For example, $PP$ stands for contraction of the operator with 
spin structure $(\gamma_5)(\gamma_5)$, $SS$ is for $(\openone )(\openone )$,
$VV$ stands for $(\gamma_\mu )(\gamma^\mu )$, and $AA$ is for
$(\gamma_\mu \gamma_5)(\gamma^\mu\gamma_5)$. The staggered flavor
is determined by the type of contraction, as explained in the
previous Appendix section. Basic contractions
are also labeled by their subscript.
The first letter indicates whether it is an ``eight'', ``eye'' or
``annihilation'' contraction, and the second is ``U'' for two, or
``F'' for one color trace.
For example: $PP_{EU}$ stands for the ``eight'' contraction
of the operator with spin-flavor structure 
$(\gamma_5\otimes\xi_5)(\gamma_5\otimes\xi_5)$ with two color traces; 
$VA_{A1F}$ stands for the ``annihilation'' contraction
of the first type, in which the derivative is taken with respect to
quark mass on the external leg (see the previous Appendix section),
the spin-flavor structure is $(\gamma_\mu\otimes\xi_5)
(\gamma^\mu\gamma_5\otimes\openone )$,
and one color trace is taken. What follows are the full 
expressions\footnote{Signs of operators $O_7$ and $O_8$ have been changed
in order to be consistent with the sign convention of Buras 
{\it et al.}~\cite{buras}.}.

``Eight'' parts:
\begin{eqnarray}
E_1^{(0)} & = &\frac{2}{3}(VV_{EF} + AA_{EF}) - \frac{1}{3}(VV_{EU}+AA_{EU}) \\
E_2^{(0)} & = &\frac{2}{3}(VV_{EU} + AA_{EU}) - \frac{1}{3}(VV_{EF}+AA_{EF}) \\
E_3^{(0)} & = & VV_{EF}+AA_{EF} \\
E_4^{(0)} & = & VV_{EU}+AA_{EU} \\
E_5^{(0)} & = & 2(PP_{EF}-SS_{EF}) \\
E_6^{(0)} & = & 2(PP_{EU}-SS_{EU}) \\
E_7^{(0)} & = & SS_{EF} - PP_{EF} +\frac{1}{2}(VV_{EU}- AA_{EU}) \\
E_8^{(0)} & = & SS_{EU} - PP_{EU} +\frac{1}{2}(VV_{EF}- AA_{EF}) \\
E_9^{(0)} & = & -E_{10}^{(0)}=\frac{1}{2} (VV_{EF}+AA_{EF}-VV_{EU}-AA_{EU}) \\
E_1^{(2)} & = & E_2^{(2)} = \frac{2}{3}E_9^{(2)} = \frac{2}{3}E_{10}^{(2)}
= \frac{1}{3} (VV_{EU}+AA_{EU}+VV_{EF}+AA_{EF}) \\
E_3^{(2)} & = &  E_4^{(2)} = E_5^{(2)} = E_6^{(2)}  = 0 \\
E_7^{(2)} & = & \frac{1}{2} (AA_{EU} - VV_{EU}) + SS_{EF}-PP_{EF} \\
E_8^{(2)} & = & \frac{1}{2} (AA_{EF} - VV_{EF}) + SS_{EU}-PP_{EU} 
\end{eqnarray}

``Eye'' parts:
\begin{eqnarray}
I_1^{(0)} & = & VV_{IU}+AA_{IU} \\
I_2^{(0)} & = & VV_{IF}+AA_{IF} \\
I_3^{(0)} & = & 3(VV_{IU}+AA_{IU}) + 2(VV_{IF}+AA_{IF}) \\
I_4^{(0)} & = & 3(VV_{IF}+AA_{IF}) + 2(VV_{IU}+AA_{IU}) \\
I_5^{(0)} & = & 3(VV_{IU}-AA_{IU}) + 4(PP_{IF} - SS_{IF}) \\
I_6^{(0)} & = & 3(VV_{IF}-AA_{IF}) + 4(PP_{IU} - SS_{IU}) \\
I_7^{(0)} & = & 2(PP_{IF} - SS_{IF}) \\
I_8^{(0)} & = & 2(PP_{IU} - SS_{IU}) \\
I_9^{(0)} & = & VV_{IF} + AA_{IF} \\
I_{10}^{(0)} & = & VV_{IU} + AA_{IU} 
\end{eqnarray}

``Annihilation'' parts are obtained by inserting the 
derivative with respect to $(m_d-m_s)$ into every propagator involving 
the strange quark:
\begin{eqnarray}
A_1^{(0)} & = & -(VA_{A1U} + AV_{A1U}) \\
A_2^{(0)} & = & -(VA_{A1F} + AV_{A1F}) \\
A_3^{(0)} & = & -3(VA_{A1U} + AV_{A1U}) - (VA_{A2U} + AV_{A2U}) \nonumber \\
& & -2 (VA_{A1F} + AV_{A1F}) - (VA_{A2F} +AV_{A2F})  \\
A_4^{(0)} & = & -3(VA_{A1F} + AV_{A1F}) - (VA_{A2F} + AV_{A2F}) \nonumber \\
& & -2 (VA_{A1U} + AV_{A1U}) - (VA_{A2U} +AV_{A2U})  \\
A_5^{(0)} & = & 3 (VA_{A1U} - AV_{A1U}) + (VA_{A2U} - AV_{A2U})
+ 2(PS_{A2F} - SP_{A2F}) \\
A_6^{(0)} & = & 3 (VA_{A1F} - AV_{A1F}) + (VA_{A2F} - AV_{A2F})
+ 2(PS_{A2U} - SP_{A2U}) \\
A_7^{(0)} & = & \frac{1}{2}(VA_{A2U} - AV_{A2U}) + (PS_{A2F} - SP_{A2F}) \\
A_8^{(0)} & = & \frac{1}{2}(VA_{A2F} - AV_{A2F}) + (PS_{A2U} - SP_{A2U}) \\
A_9^{(0)} & = & VA_{A1F} + AV_{A1F} + \frac{1}{2} (VA_{A2U}+AV_{A2U}+VA_{A2F}+AV_{A2F}) \\ 
A_{10}^{(0)} & = & VA_{A1U} + AV_{A1U} + \frac{1}{2} (VA_{A2F}+AV_{A2F}+VA_{A2U}+AV_{A2U}) \\ 
\end{eqnarray}
Of course, ``eye'' and ``annihilation'' contractions are not present 
in $I=2$ operators.


\pagebreak

\centerline{\large TABLES\\} 
\begin{table}[!h]
\caption{$\varepsilon '/\varepsilon$ in units of $10^{-4}$ for $Q_1$
ensemble, computed in three ways: (M1) $\Re A_0$ and $\Re A_2$ 
values entering the expression for $\varepsilon '/\varepsilon$
are taken from experiment; (M2) $\omega$ amplitude ratio is taken from
experiment, while $\Re A_0$ amplitude is from our simulation;
(M3) both $\Re A_0$ and $\Re A_2$ are from our simulation. 
Partially-nonperturbative matching have been used to obtain 
the results. In all perturbative corrections
we have used one-loop coefficients computed for 
gauge-noninvariant operators.
The first error is statistical (obtained by combining the individual 
errors in matrix elements by jackknife).
The second error is the diagonal operator matching error due to uncertainty 
in the determination of $Z_P$ and $Z_S$. 
In order to estimate the non-diagonal matching error we compare
two renormalization procedures: using strictly one-loop 
non-diagonal corrections (denoted ``(1l.)''), and resumming part of 
higher-order corrections in non-diagonal mixing by using 
non-perturbative renormalization factors $Z_P$ and $Z_S$ (as explained in 
Section~\ref{sec:ansatz}). The latter method is denoted ``(p.r.)''. 
Some other parameters used in obtaining
these results are: $\Im \lambda_t = 1.5\cdot 10^{-4}$, $m_t=170$ GeV,
$m_b=4.5$ GeV, $m_c=1.3$ GeV, $\Omega_{\eta +\eta '} = 0.25$,
$\alpha^{(n_f=0)}_{\overline{\mathrm MS}} (2\quad {\mathrm GeV}) = 0.195$
(the latter is based on setting the lattice scale by $\rho$ meson mass).
Short distance coefficients were obtained by two-loop running
using the anomalous dimension and threshold matrices computed by
Buras {\it et al.} [12].}
\label{tab:epsp1}
\begin{tabular}{l|ccc}
\hline\hline
Quark mass & 0.01 & 0.02 & 0.03  \\
\tableline
M1 (p.r.) &
$-58.1  \pm 2.1   \pm 10.6 $ & 
$-26.3   \pm 0.8   \pm 8.9 $ & 
$-16.1 \pm 0.4 \pm 8.0 $ \\ 
M1 (1l.) &
$-52.3 \pm 2.2 \pm 10 $ &
$-22.0 \pm 0.8 \pm 8.3 $ &
$-12.2 \pm 0.5 \pm 6.9 $ \\
M2 (p.r.) &
$-38.6 \pm 2.1 \pm 6.0 $ &   
$-18.7 \pm 0.3 \pm 7.0 $ &  
$-11.7 \pm 0.2 \pm 6.0 $ \\ 
M2 (1l.) &
$-45.4 \pm 3.5 \pm 8.6 $ & 
$-18.8 \pm 0.4 \pm 7.0 $ &
$-10.3 \pm 0.3 \pm 6.0 $ \\
M3 (p.r.) &     
$-97   \pm 14  \pm 13 $ &   
$-81   \pm 4   \pm 23 $ &    
$-79   \pm 2   \pm 27 $ \\  
M3 (1l.) &
$-142 \pm 28 \pm 29 $ &
$-88 \pm 5 \pm 35 $ &
$-75 \pm 2 \pm 39 $ \\
\hline
\hline
Quark mass & 0.04 & 0.05 &\\
\tableline
M1 (p.r.) &
$-8.0 \pm 0.9 \pm 7.2  $ &        
$-4.4  \pm 0.9 \pm 7.2 $ & \\   
M1 (1l.) &
$-4.2 \pm 1.1 \pm 6.5 $ &
$-1.2 \pm 1.0 \pm 6.6 $ \\
M2 (p.r.) &
$-6.1  \pm 0.5 \pm 5.3  $ &   
$-3.1  \pm 0.5 \pm 4.9  $ & \\  
M2 (1l.) &
$-3.7 \pm 0.8 \pm 5.8 $ &
$-0.9 \pm 0.8 \pm 5.2 $ \\
M3 (p.r.) &
$-81   \pm 5   \pm 38  $ &   
$-74   \pm 4   \pm 38  $& \\ 
M3 (1l.) &
$-64 \pm 4 \pm 52 $ &
$-55 \pm 5 \pm 51$ \\
\hline\hline
\end{tabular}
\end{table}

\begin{table}[tbph]
\caption{$\varepsilon '/\varepsilon$ results for $Q_3$ ensemble 
($\beta=6.2$). Everything else is the same as in Table~\ref{tab:epsp1}.}
\label{tab:epsp2}
\begin{center}
\begin{tabular}{l|cc}
\hline\hline
Quark mass & 0.010 & 0.015 \\
\tableline
M1 (p.r.) &
$  -109\pm   15  \pm   84 $ &     
$  -64 \pm   7   \pm   63 $ \\    
M1 (1l.) &
$ -95 \pm 16 \pm 74 $&
$-52 \pm 8 \pm 56$ \\
M2 (p.r.) &
$  -39 \pm  9   \pm     22 $ &    
$  -22 \pm  2   \pm     18 $ \\   
M2 (1l.) &
$ -48 \pm 18 \pm 33$ &
$-23 \pm 3 \pm 23 $ \\
M3 (p.r.) &
$-30.0 \pm 17 \pm 23.2 $ &      
$-21.1 \pm 2.3 \pm   20.7 $ \\  
M3 (1l.) &
$-48 \pm 42 \pm 39$ &
$-24 \pm 7 \pm 28$ \\ 
\hline
\hline
Quark mass & 0.020 & 0.030  \\
\tableline
M1 (p.r.) &
$  -36 \pm   4   \pm   60 $ &      
$  -19 \pm   2   \pm   38 $ \\     
M1 (1l.) &
$-27 \pm 5 \pm 54 $ &
$-14 \pm 2 \pm 34 $ \\
M2 (p.r.) &
$  -13.7 \pm  0.6 \pm   20 $ &     
$  -9.5 \pm  0.6 \pm    17 $ \\    
M2 (1l.) &
$-12.4 \pm 1.1 \pm 23.3 $ &
$-8.1 \pm 0.9 \pm 18.5 $ \\ 
M3 (p.r.) &
$-15.9 \pm 1.8  \pm   26.5 $ &    
$-10.8 \pm 1.1 \pm   21.6 $ \\    
M3 (1l.) &
$-14.7 \pm 1.4 \pm 31.1 $ &
$-9.6 \pm 0.8 \pm 18.4 $ \\
\hline\hline
\end{tabular}
\end{center}
\end{table}

\end{document}